\documentclass[fleqn,usenatbib,letterpaper,oneside]{mnrasletter}

\usepackage[totalwidth=480pt,totalheight=680pt,layoutvoffset=0.5cm]{geometry}

\usepackage{newtxtext,newtxmath}

\usepackage[T1]{fontenc}
\usepackage{ae,aecompl}

\usepackage{graphicx}	
\usepackage{amsmath}	
\usepackage{amssymb}	
\usepackage{breqn}
\usepackage{multirow}
\usepackage{rotating}\usepackage{adjustbox}

\title[Planets and debris discs]{No significant correlation between radial velocity planet presence and debris disc properties}

\author[B. Yelverton et al.]{
Ben Yelverton$^{1}$\thanks{E-mail: bmy21@cam.ac.uk},
Grant M. Kennedy$^{2,3}$ and
Kate Y. L. Su$^{4}$ 
\\
$^{1}$Institute of Astronomy, University of Cambridge, Madingley Road, Cambridge CB3 0HA, UK\\
$^{2}$Department of Physics, University of Warwick, Gibbet Hill Road, Coventry CV4 7AL, UK\\
$^{3}$Centre for Exoplanets and Habitability, University of Warwick, Gibbet Hill Road, Coventry CV4 7AL, UK\\
$^{4}$Steward Observatory, University of Arizona, 933 N Cherry Avenue, Tucson, AZ 85721, USA
}

\date{Accepted XXX. Received YYY; in original form ZZZ}

\pubyear{2020}

\begin{document}
\label{firstpage}
\pagerange{\pageref{firstpage}--\pageref{lastpage}}
\maketitle

\begin{abstract}

We investigate whether the tentative correlation between planets and debris discs which has been previously identified can be confirmed at high significance. We compile a sample of 201 stars with known planets and existing far infrared observations. The sample is larger than those studied previously since we include targets from an unpublished \textit{Herschel} survey of planet hosts. We use spectral energy distribution modelling to characterise Kuiper belt analogue debris discs within the sample, then compare the properties of the discs against a control sample of 294 stars without known planets. Survival analysis suggests that there is a significant ($p\sim 0.002$) difference between the disc fractional luminosity distributions of the two samples. However, this is largely a result of the fact that the control sample contains a higher proportion of close binaries and of later-type stars; both of these factors are known to reduce disc detection rates. Considering only Sun-like stars without close binary companions in each sample greatly reduces the significance of the difference ($p\sim 0.3$). We also find no evidence for a difference in the disc fractional luminosities of stars hosting planets more or less massive than Saturn ($p\sim 0.9$). Finally, we find that the planet hosts have cooler discs than the control stars, but this is likely a detection bias, since the warmest discs in the control sample are also the faintest, and would thus be undetectable around the more distant planet hosts. Considering only discs in each sample that could have been detected around a typical planet host, we find $p\sim 0.07$ for the temperatures.

\end{abstract}




\section{Introduction}
\label{sec:intro}

A significant fraction of main-sequence stars show excess emission at infrared wavelengths, attributed to thermal emission from dust in a circumstellar debris disc (see e.g. \citealt{Wyatt08_Review}; \citealt{Krivov10_Review}; \citealt{Matthews14_DiscReview}; \citealt{Hughes18_Review}). The inferred dust should have a short lifetime due to destructive collisions and radiation forces, and this fact has led to the conclusion that a population of undetected planetesimals must be acting as a source of new dust through its own collisional evolution (\citealt{Backman93_VegaPhenom}; \citealt{Dominik03_VegaTheory}). Since planetesimals are both an important component of debris discs and a key building block in the formation of planets (\citealt{Johansen14_Planetesimals}), one might expect the properties of debris discs to be correlated with the properties of any planets in a given system. Previous work has suggested that such correlations may indeed exist, but has remained somewhat inconclusive (see below). A full understanding of these correlations could aid our understanding of how planetary systems form and evolve, and perhaps enable more efficient searches for planets.

Various theoretical arguments have been made for the possible existence of a planet-disc correlation. \citet{Wyatt07_Metallicity} considered a model in which planets only form in a protoplanetary disc if the mass of the disc exceeds some critical value, and in which the protoplanetary disc mass also determines the initial mass of the debris disc which it ultimately forms. They predicted that debris discs detected around planet-hosting stars should have greater fractional luminosities (i.e. ratios of disc to stellar luminosity) than those in systems with no planets, since within their model planets form from high-mass protoplanetary discs, which also go on to produce a greater mass of planetesimals. In addition, it may be expected that such a correlation is dependent on planetary mass. \citeauthor{Raymond11_TerrestrialPlts1} (\citeyear{Raymond11_TerrestrialPlts1}, \citeyear{Raymond12_TerrestrialPlts2}) studied simulations of systems with an inner disc of planetesimals and planetary embryos, three giant planets at semi-major axes $\sim$5--10~au, and an outer debris disc. They found that the giant planets tend to either leave both the inner and outer discs largely unperturbed or to perturb them both simultaneously, such that the systems in which the outer disc can survive on long time-scales are also those in which terrestrial planet formation in the inner disc can proceed unimpeded. This led to the prediction that debris discs should be correlated with terrestrial planets, but anti-correlated with eccentric giant planets. \citet{Wyatt12_LowMassPlts} came to a similar conclusion by considering two planet formation scenarios. If planets form at relatively large separations and then migrate inwards, then more massive planets will cause a larger fraction of the planetesimals they encounter during the migration to be ejected, so that low-mass planets are more conducive to the survival of outer planetesimal belts. If instead planets form in situ at relatively close separations, then low-mass planets could scatter planetesimals to large distances where they may form an outer belt, whereas more massive planets are more likely to simply eject planetesimals. Thus, in either scenario, it is plausible that systems with only low-mass planets are more likely to host a debris disc.

There has also been much observational effort to identify any correlation between planets and debris discs. \citet{Bryden09_PlanetDisc} studied a sample of 146 known radial velocity (RV) planet-hosting stars along with a sample of 165 stars not known to host planets, using data from the Multiband Imaging Photometer for \textit{Spitzer} (MIPS; \citealt{Rieke04_MIPS}) to compare the properties of the debris discs detected in both samples. Their results suggested that the planet-hosting stars do tend to have brighter debris discs than those not known to host planets by a factor of $\sim$2, though the difference was not statistically significant, with a confidence level of only 77\% that the fractional luminosities of the discs in the two samples were drawn from different distributions. Similarly, \citet{Kospal09_PlanetDisc} found using MIPS observations of 150 planet hosts (largely the same as those studied by \citealt{Bryden09_PlanetDisc}) and 118 stars with no detected planets that the debris discs around the planet hosts tended to be brighter, but not in a statistically significant way, with just an 87\% probability that the distributions of the ratio of observed to predicted flux at 70~$\mu$m were different for the two samples. More recently, \citet{Matthews14_DiscReview} have stated in their review article that an analysis of \textit{Herschel} Photodetector Array Camera and Spectrometer (PACS; \citealt{Poglitsch10_PACS}) data from a survey of 69 planet-bearing stars finds that planet hosts have brighter debris discs than stars without known planets at greater than $3\sigma$ confidence (see their Fig.~4). However, the details of the analysis behind this claim -- for example, how the stellar samples were constructed, how the presence of an infrared excess was established, and which statistical methods were used  -- have not been published.

\citet{Wyatt12_LowMassPlts} found some evidence for a correlation between debris discs and low-mass planets through analysis of a sample of the nearest 60 G-type stars. They found that 4/6 of the stars in this sample with planets less massive than Saturn showed excess infrared emission indicative of a debris disc, in contrast with 0/5 of those with more massive planets. They demonstrated that despite the small number statistics, there is only a chance of a few per cent of detecting at least four discs out of a sample of six given the expected detection rate of $\sim$15--20\% for Sun-like stars (\citealt{Trilling08_SunLikeStars}; \citealt{Eiroa13_DUNES}; \citealt{Sierchio14_DiscDecay}; \citealt{Sibthorpe18_DEBRIS}). \citet{Marshall14_Correlations} arrived at a similar result using PACS observations of a volume-limited sample of 37 exoplanet host stars: they concluded that 6/11 stars for which the most massive planet was less than 30$M_{\oplus}$ host a debris disc, compared with only 5/26 of those with a planet above this limit. However, \citet{MoroMartin15_PlanetsAndKBs} later studied a sample of stars older than 1~Gyr and with no binary companion within 100~au, compiled from the DEBRIS (\citealt{Matthews10_DEBRIS}) and DUNES (\citealt{Eiroa10_DUNES}) \textit{Herschel} surveys (and containing both the \citealt{Wyatt12_LowMassPlts} and \citealt{Marshall14_Correlations} subsamples), concluding that there was no significant correlation between debris discs and either low or high-mass planets. However, this finding could simply reflect the fact that the number of planet hosts in the DEBRIS and DUNES samples is low (a few tens), since these surveys did not specifically target stars with planets. Finally, while most previous work has focused on close-in planets discovered through the radial velocity technique, \citet{Meshkat17_DIPlanets} have recently suggested a tentative (88\% confidence) correlation between debris discs and directly imaged giant planets beyond $\sim$10~au. 

It is clear, then, that the question of whether planets and debris discs are correlated is not a solved problem. However, it appears from \citet{Matthews14_DiscReview} that there now exists far infrared data -- which allows the detection and characterisation of Kuiper belt analogue debris discs -- for sufficiently many planet hosts to allow the identification of a statistically significant difference between the disc properties of planet-bearing and planet-free stars. This fact motivates a revisitation of the issue of planet-disc correlation, and this is the issue that we address in this paper. To do so, we compile two samples of stars with existing PACS and/or MIPS observations: one in which the stars host at least one planet discovered using the RV method, and one with no known planets. In section~\ref{sec:samples} we explain how the stars were chosen, and summarise the stellar properties of both samples. We model the spectral energy distributions (SEDs) of the stars to identify and characterise their Kuiper belt analogue debris discs; this process is detailed in section~\ref{sec:SED_modelling}. In section~\ref{sec:results}, we present the disc properties derived from the SED modelling, then proceed to compare the fractional luminosity and temperature distributions of the discs in the two samples statistically, and consider what our results may imply for the formation and subsequent migration of planets. We summarise our conclusions in section~\ref{sec:conclusions}.

\section{The samples}
\label{sec:samples}

In this paper we study two samples of stars, distinguished by the presence or absence of known planets. This section describes how the samples were compiled, before comparing their stellar properties and considering how the differences in these properties may bias our results.

\subsection{Planet host sample selection}
\label{sec:planet_sample}

Our sample of planet-hosting stars is a subset of the NASA Exoplanet Archive\footnote{Downloaded on October 14, 2019} (NEA; \citealt{Akeson13_NASAArchive}), resulting from the application of the cuts listed below. There are 201 stars in the sample, and their properties are summarised in Table~\ref{tab:planetsample}. Note that in this Table, the stellar effective temperatures $T_{\mathrm{eff}}$ and luminosities $L$ are derived from our SED modelling (outlined in section~\ref{sec:SED_modelling}). $T_{\mathrm{eff}}$ is typically precise to $\sim$100~K, while $L$ is typically precise to $\sim$5\% (but can be more uncertain when the stellar distance is known less precisely than $\sim$5\%).

\begin{table*}
\begin{tabular}{lrrrrrr}
\hline
Name & $d$ / pc & $T_{\mathrm{eff}}$ / K & $L / L_{\odot}$ & $a_{\mathrm{binary}}$ / au & $M\sin(i)$ / $M_{\mathrm{J}}$ & $a_{\mathrm{pl}}$ / au \\
\hline

HD 142&	26.2&	6200&	3.0E+00&	1.0E+02&	5.3E+00&	6.8E+00 \\
HD 1237&	17.6&	5460&	6.5E-01&	6.9E+01&	3.4E+00&	4.9E-01 \\
HD 1326&	3.6&	3444&	2.1E-02&	2.3E+01&	1.1E-01& 	5.4E+00 \\
HD 1461&	23.5&	5736&	1.2E+00&	&2.0E-02&	6.3E-02 \\
HD 3651&	11.1&	5240&	5.3E-01&	&2.3E-01&	3.0E-01 \\
HD 4203&	81.6&	5577&	2.0E+00&	&2.2E+00&	1.2E+00 \\
HD 4208&	34.2&	5708&	8.0E-01&		&8.1E-01&	1.7E+00 \\
HD 4308&	22.0&	5676&	1.0E+00&		&5.0E-02&	1.2E-01 \\
HD 6434&	42.4&	5789&	1.3E+00&		&4.9E-01&	1.4E-01 \\
HD 7924&	16.8&	5238&	3.9E-01&		&2.5E-02&	1.1E-01 \\

\multicolumn{7}{c}{$\cdots$} \\

\hline
\end{tabular}
\caption{List of stars in our planet host sample, together with their distances (from the NASA Exoplanet Archive), effective temperatures and luminosities (from our SED modelling in section~\ref{sec:SED_modelling}), closest relevant binary separation (from the Washington Double Star Catalog; see section~\ref{sec:planet_sample} for details), and $M\sin(i)$ values and semi-major axes of the most massive planet in each system (from NEA). Typical uncertainties on $d$, $T_{\mathrm{eff}}$ and $L$ are $\sim$0.1~pc, $\sim$100~K and $\sim$5\% respectively. Here we show only the first ten lines; a full machine-readable version can be obtained online.}
\label{tab:planetsample}
\end{table*}

\begin{itemize}
    
    \item Observations at far infrared wavelengths are necessary for the characterisation of cool debris discs (whose temperatures typically lie in the range $\sim$20--200~K), which is the aim of this paper. Therefore, we select only stars which have been observed by MIPS at 70~$\mu$m and/or by PACS at 70 or 100~$\mu$m, determined by cross-matching NEA with the \textit{Spitzer} Heritage Archive\footnote{\url{https://sha.ipac.caltech.edu/applications/Spitzer/SHA/}} and the \textit{Herschel} Science Archive\footnote{\url{http://archives.esac.esa.int/hsa/whsa/}}. As a result, our sample will necessarily be dominated by the targets of previous far infrared surveys of planet hosts -- that is, the stars observed by MIPS and discussed by \citet{Bryden09_PlanetDisc}, combined with the unpublished PACS survey \texttt{OT1_gbryden_1} (\citealt{Bryden10_Proposal}) which was used to create Fig.~4 of \citet{Matthews14_DiscReview}, and the PACS survey of late K and M-type planet hosts discussed by \citet{Kennedy18_MStarDiscs} -- with the addition of stars targeted by other surveys but which by chance host detected planets. The planet host surveys have focused on systems with planets discovered using Doppler spectroscopy, also known as the RV method. This is largely because planets discovered through the transit method tend to be much further away than RV planets, as the specific orientation required for transits to be observed means that they are rarer (e.g. from NEA, 624/776 planets discovered through RV are within 100~pc, compared with only 139/3123 of those discovered through their transits). Far infrared surveys of transiting planet hosts would thus have relatively poor sensitivity. There is only a single system with a transit-discovered planet that survives all of our cuts: GJ~1214 (from the \citealt{Kennedy18_MStarDiscs} survey)\footnote{HD~39091 (pi Men) hosts both an RV planet and a transiting planet; the former was discovered first, and is the more massive of the two.}.
    
    \item We exclude systems with planets discovered through direct imaging (DI), firstly because DI is sensitive to a very different population of planets than RV (typically only planets of at least several Jupiter masses at tens of au are detectable; e.g. \citealt{Apai08_DISurvey}; \citealt{Matthews18_TwoBeltDI}), and secondly because the targets of DI surveys will be systematically younger than those of RV surveys, since younger planets are brighter and thus easier to detect.
    
     \item More distant systems will have a higher detection threshold for circumstellar dust, so we apply a distance cut, including only systems closer than 100~pc (using the distances listed in NEA). Taking a closer cutoff distance would improve the sample's typical sensitivity to disc emission, but result in a smaller sample with limited statistical power.  
    
    \item We exclude stars which are clearly giants or subgiants based on their positions on a Hertzsprung-Russell (H-R) diagram using effective temperatures and luminosities from our SED modelling (see section~\ref{sec:SED_modelling}). This is to allow a fairer comparison with our control sample, which is composed only of FGKM-type main-sequence stars (see subsection~\ref{sec:control_sample}). Having left the main sequence, giants and subgiants will either be older than the stars in our control sample, or have evolved from a main-sequence star of earlier spectral type. In either case, leaving such stars in the planet host sample would affect our results when comparing the disc populations of the two samples, as it has been well established that debris discs are detected more frequently (and with greater fractional luminosity) around younger stars and those with earlier spectral types (e.g. \citealt{Su06_AStarEvolution}; \citealt{Hillenbrand08_FEPS}; \citealt{Trilling08_SunLikeStars}; \citealt{Urban12_IncidenceOfDiscs}; \citealt{Eiroa13_DUNES}; \citealt{Sierchio14_DiscDecay}; \citealt{Thureau14_HerschelA}; \citealt{Sibthorpe18_DEBRIS}). The age dependence of the detection rate follows from collisional depletion (e.g. \citealt{Wyatt07_Transience}; \citealt{Lohne08_CollEvol}), with the spectral type dependence being a related phenomenon, since later-type stars tend to be older due to their longer main-sequence lifetimes; in addition, sensitivity to dust is a strong function of spectral type, with discs more difficult to detect around later-type stars (see e.g. Fig.~2 of \citealt{Matthews14_DiscReview}).
    
    \item Finally, we examined the far infrared images of all candidates remaining after the above cuts, and discarded the following five systems based on apparent contamination either by background sources or infrared cirrus, which could lead to false positive excess detections: HD~19994, HD~46375, HD~70642, HD~190360 and HIP~79431.
\end{itemize}

In order to fully understand the resulting sample and correctly interpret our results, it is necessary to know the multiplicity properties of the stars. This is important as it has been found that debris discs are detected significantly less often in binaries with separations closer than $\sim$100~au than in wider binaries or around single stars (\citealt{Trilling07_BinaryDebris}; \citealt{Rodriguez12_BinaryDebris}; \citealt{Rodriguez15_BinaryDebris}; \citealt{Yelverton19_BinaryDebris}). As multiplicity information is not included in NEA, we cross-match our sample with the Washington Double Star Catalog\footnote{Accessed on October 23, 2019} (WDS; \citealt{Mason01_WDS}). For systems appearing in WDS, we calculate the average of the angular separations at the two epochs listed, then convert this to a projected physical separation using distances from NEA. In cases where there are three or more components (and thus more than one separation) listed, we record the \textit{closest} relevant separation in Table~\ref{tab:planetsample}, since wide companions do not have a significant influence on disc detection rate (\citealt{Yelverton19_BinaryDebris}). The stars in the planet host sample are largely primaries, so for most systems we ignore separations not involving the primary; e.g. in a triple system with an A component and a distant B component which itself is a close binary, the BaBb separation would be the closest, but as this is irrelevant to a circumprimary disc we would record the AB separation. The two exceptions to this rule are HD~178911~B and HD~186427 (16~Cyg~B), which are both tertiary components of triple systems with a close AaAb separation and a wider AB separation (see the Multiple Star Catalog, MSC; \citealt{Tokovinin18_MSC}). As the stars we are interested in are the B components, the AaAb separation is not relevant in these cases, so we instead record the wider AB separation\footnote{HD~162004 (Psi$^1$~Dra~B) is the B component of a system with similar architecture (see MSC). However, WDS does not list the AaAb separation.}. 

As WDS is a catalogue of visual binaries, we also used the Ninth Catalogue of Spectroscopic Binary Orbits\footnote{Accessed on October 23, 2019} (SB9; \citealt{Pourbaix04_SB9}) to check for any very close binaries. The only star listed in SB9 is HD~114762, with a period of 83.9~d\footnote{HD~114762 also has a much wider stellar companion listed in WDS, with a separation of $\sim$130~au.} (from \citealt{Mazeh96_HD114762}); this is the same period as the planet listed in NEA with $M\sin(i)\approx 11M_{\mathrm{J}}$, and the entries in the two catalogues in fact refer to the same object. The nature of HD~114762~b has been unclear since its discovery by \citet{Latham89_HD114762}, with recent analysis of \textit{Gaia} astrometry suggesting that it is on an almost face-on orbit, implying that it may in fact be an M-type star rather than a planet (\citealt{Kiefer09_HD114762}). Note that we do not exclude HD~114762 from the outset based on the results of \citet{Kiefer09_HD114762}; it would not be fair to do so since there may be other stars in the planet host sample whose companions could be shown to be stellar if a similar analysis were performed. However, this system highlights a general issue with RV planet candidates with large $M\sin(i)$: the fact that their inclinations are usually unknown means that they could be planets, brown dwarfs, or even stars. The inability to distinguish substellar from stellar companions purely through RV could be problematic in the sense that, as discussed above, close binaries are known to adversely affect the incidence of debris discs. Thus, if planets are positively correlated with debris discs but the planet host sample contains some systems which are actually close binaries, then the correlation will appear less strong than it really is. In section~\ref{sec:stellarcompanions}, we consider how this issue might be addressed by obtaining a second estimate of the masses of companions using astrometry.

\begin{figure}
	\centering
    \hspace{-0.5cm}
	\includegraphics[width=0.5\textwidth]{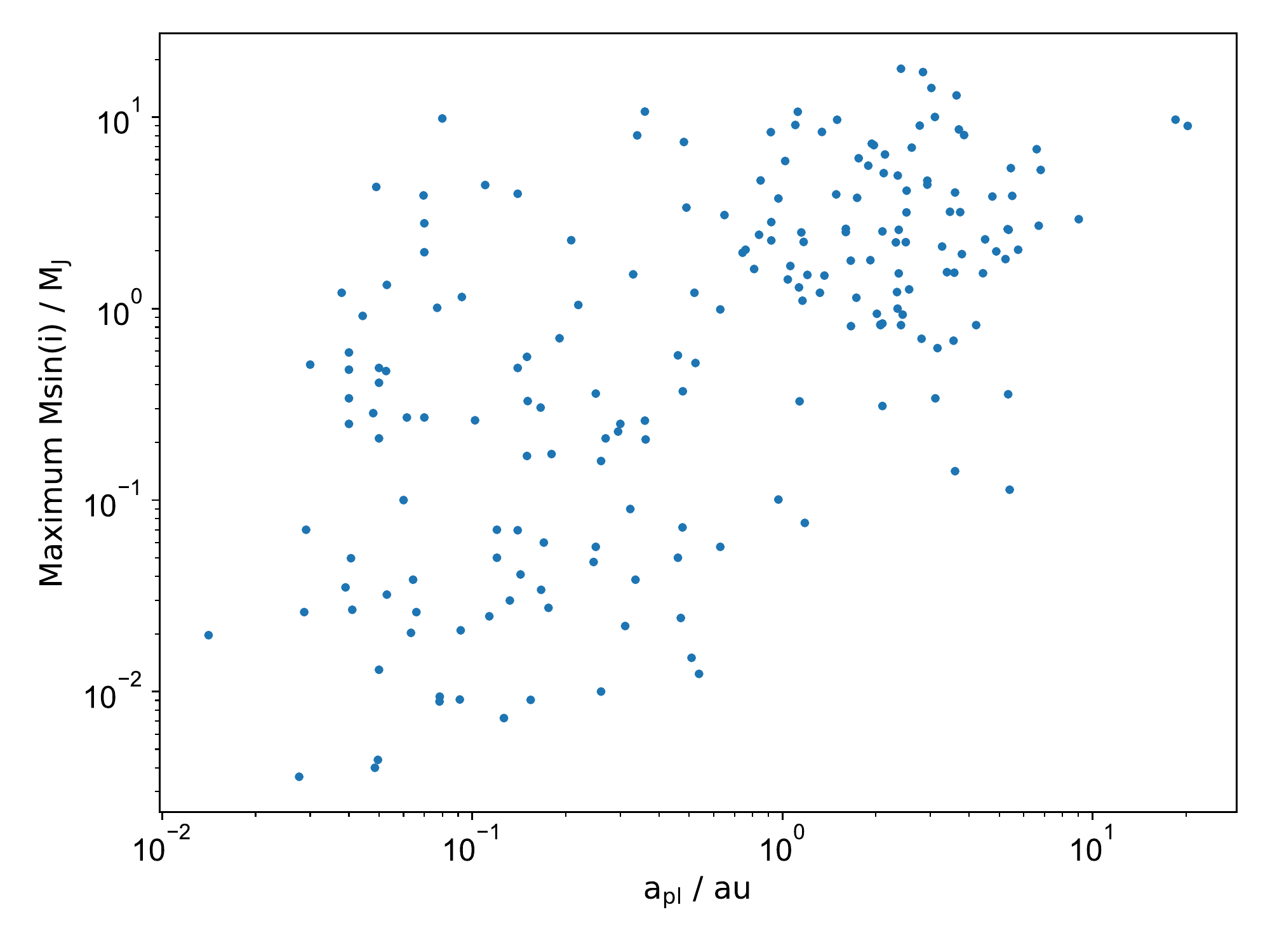}
	\caption{Summary of the properties of the planets in our planet host sample, showing $M\sin(i)$ values and semi-major axes. In the case of multi-planet systems, we plot only the planet with the largest $M\sin(i)$. For GJ~1214 the value plotted on the vertical axis is simply the planetary mass, as its planet was discovered via the transit method rather than the radial velocity method.}
	\label{fig:plt_properties}
\end{figure}

The $M\sin(i)$ values and semi-major axes $a_{\mathrm{pl}}$ of the planets orbiting the stars in our sample are shown in Fig.~\ref{fig:plt_properties}. For the 62 stars with multiple planets, we plot (and show in Table~\ref{tab:planetsample}) only the parameters of the planet with the largest $M\sin(i)$; we are primarily interested in the most massive planet in a given system, since based on theoretical arguments, the likelihood of detecting a debris disc may be determined by the presence or absence of giant planets (see references in section~\ref{sec:intro}). The fact that the sample is composed of RV-discovered planet hosts in all cases but one explains the distribution seen in Fig.~\ref{fig:plt_properties}; the detection limit in mass rapidly increases beyond a few au, where the observations span less than one orbital period. A sample of transiting planets would have a much smaller typical semi-major axis, since close-in planets have a much greater transit probability (e.g. \citealt{Borucki84_TransitMethod}); this explains why the only transit-discovered planet in the sample, GJ~1214~b, has the smallest semi-major axis, at 0.014~au. We choose not to exclude GJ~1214 since the mass and semi-major axis of its planet are such that it would in principle be discoverable by RV, and thus it does not belong to a fundamentally different population of planets than the rest of the sample (unlike, for example, typical directly imaged planets, which we do exclude).

\subsection{Control sample selection}
\label{sec:control_sample}

Our control sample is drawn from the DEBRIS sample, which is a volume-limited sample of nearby AFGKM main-sequence stars whose selection is detailed in  \citet{Phillips10_DEBRISTargets}. The stars we select are all within $\sim$25~pc (see Fig.~\ref{fig:sample_properties}). All systems in the sample have been observed with PACS at 70 and/or 100~$\mu$m, mostly through the DEBRIS survey (\citealt{Matthews10_DEBRIS}) and some through the DUNES survey (\citealt{Eiroa10_DUNES}). The control sample results from applying the cuts listed below, and contains 294 stars, whose properties are summarised in Table~\ref{tab:controlsample}. 

\begin{table}
\centering
\begin{tabular}{lrrrr}
\hline
Name & $d$ / pc & $T_{\mathrm{eff}}$ / K & $L / L_{\odot}$ & $a_{\mathrm{binary}}$ / au \\
\hline 
HD 166&	13.7&	5513&	6.3E-01&	\\
HD 693&	18.7&	6196&	3.0E+00&    \\
HD 739&	21.3&	6481&	3.0E+00&	\\
HD 1581&	8.6&	5965&	1.3E+00&	\\
HD 1835&	20.9&	5841&	1.1E+00&	\\
HD 3443&	15.4&	5449&	1.2E+00&	8.9E+00 \\
HD 4391&	15.2&	5781&	9.6E-01&	2.5E+02 \\
HD 4628&	7.4&	4993&	2.9E-01&	\\
HD 4676&	23.5&	6232&	4.3E+00&	1.3E-01 \\
HD 4747&	18.7&	5364&	4.5E-01&	7.2E+00 \\

\multicolumn{5}{c}{$\cdots$} \\
\hline
\end{tabular}
\caption{List of stars in our control sample, together with their distances (from \citealt{Phillips10_DEBRISTargets}), effective temperatures and luminosities (from our SED modelling in section~\ref{sec:SED_modelling}), and closest relevant binary separation (from \citealt{Rodriguez15_BinaryDebris}). Typical uncertainties on $d$, $T_{\mathrm{eff}}$ and $L$ are $\sim$0.1~pc, $\sim$100~K and $\sim$5\% respectively. Here we show only the first ten lines; a full machine-readable version can be obtained online.}
\label{tab:controlsample}
\end{table}

\begin{itemize}
    \item We retain only stars with $T_{\mathrm{eff}}<6700~\mathrm{K}$, where $T_{\mathrm{eff}}$ is the effective stellar temperature we obtain from SED modelling (see section~\ref{sec:SED_modelling}). This makes the control sample more comparable with the planet host sample, which does not contain any stars with earlier spectral type than mid-F -- a result of the fact that early-type stars are unfavourable targets for RV observations, since they have fewer spectral lines (as their higher temperatures cause a greater degree of ionisation), and the lines that they do have are broader (as they are typically rapidly rotating). As early-type stars have a higher rate of debris disc detection than late-type stars (see references in section~\ref{sec:planet_sample}), without this cut the control sample would be biased towards a higher disc detection rate.
    
    \item In the case of multiple star systems, we include only the primary, since the planet host sample is dominated by primaries. Note in particular that this results in the exclusion of Fomalhaut B and C, the latter of which is known to host a debris disc (\citealt{Kennedy14_FomalhautC}).
    
    \item We exclude systems which are known to host planets, and which are therefore already in the planet host sample (other than the two DEBRIS stars with directly imaged planets, HD~115383 and HD~206860, which we nonetheless exclude from the control sample). This was done by cross-matching the DEBRIS sample with NEA and discarding any systems which appear therein. Note that the remaining systems may still in fact host planets, either because they have simply not been targeted by planet surveys or because their planets are below the detection limits. It would be ideal to use a sample of stars which have \textit{all} been observed in RV and had no planets detected; such a sample could be obtained from e.g. \citet{Howard16_PltNonDetection}, but since \textit{Herschel} ceased observing in 2013, these stars will not generally have been observed in the far infrared.
    
\end{itemize}

Note that unlike for the planet host sample, we do not need to discard any systems on the basis of contamination. Systems predicted to have high levels of infrared cirrus contamination were not included as targets for the DEBRIS survey (\citealt{Phillips10_DEBRISTargets}). For the FGK stars, we use photometry from \citet{Sibthorpe18_DEBRIS}; as discussed in that paper, two systems (HD~20807 and HD~88230) do appear to be contaminated by background sources, but this was accounted for in the photometry by fitting multiple point sources near the star.

As with the planet host sample, we also need to know about the multiplicity of the control stars. For this sample it is not necessary to cross-match with WDS, as multiplicity information on all of the DEBRIS stars is readily available from \citet{Rodriguez15_BinaryDebris}. Thus, we simply use Tables 1 and 2 of that paper to record whether or not each system has any companions, and if so, the closest separation involving the primary (shown in Table~\ref{tab:controlsample}).

\subsection{Comparison of stellar properties}
\label{sec:sample_properties}

\begin{figure}
	\centering
    \hspace{-0.5cm}
	\includegraphics[width=0.5\textwidth]{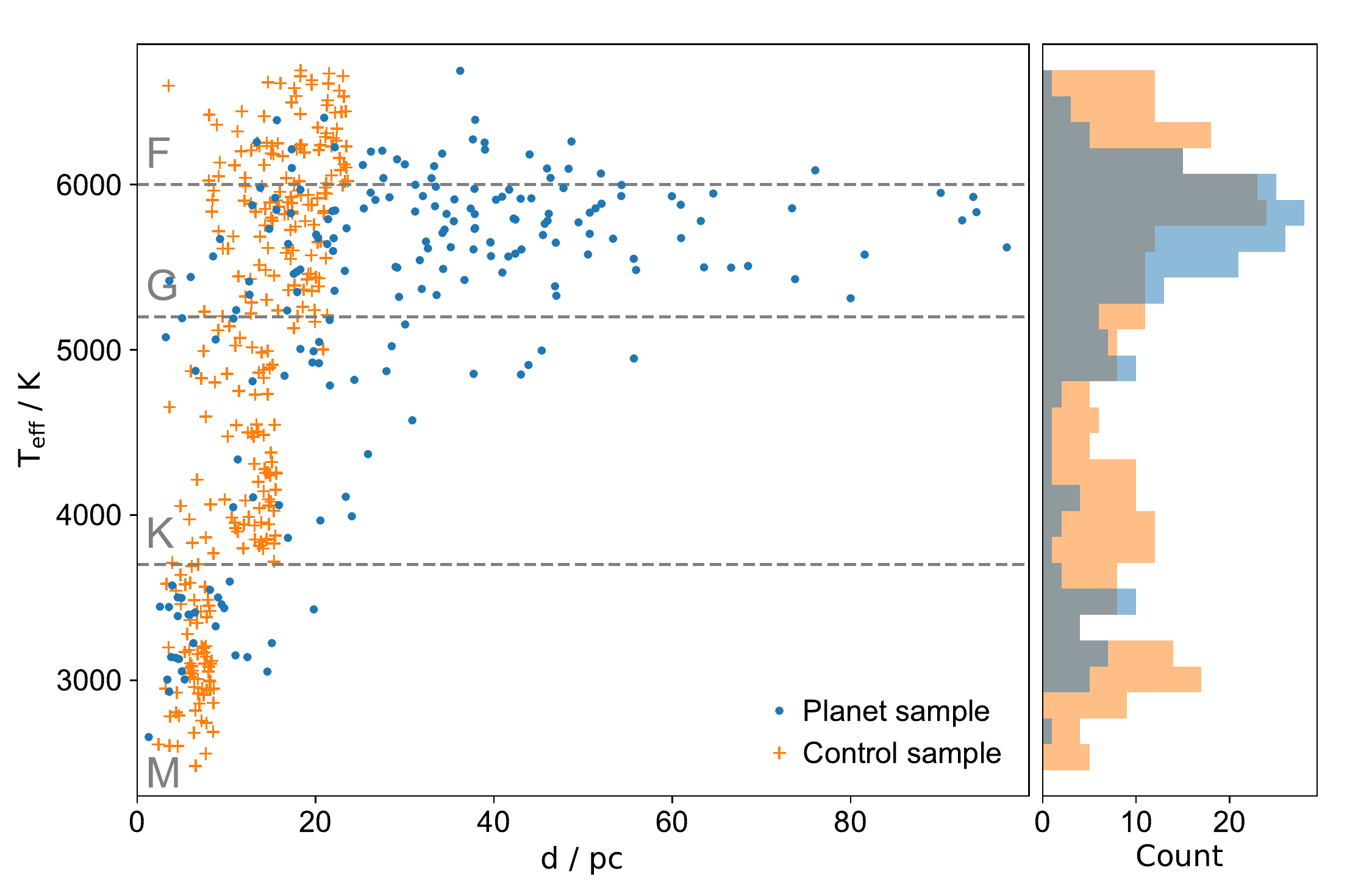}
	\caption{Effective stellar temperature from SED modelling (see section~\ref{sec:SED_modelling}) versus distance for both our planet host sample and control sample. The histogram shows the distributions of effective temperature, and the dashed lines show approximate boundaries between spectral types.}
	\label{fig:sample_properties}
\end{figure}

In Fig.~\ref{fig:sample_properties} we show the distances and effective temperatures of both samples. The distances of the planet sample are taken from NEA, while those of the control sample are from \citet{Phillips10_DEBRISTargets}; all temperatures are derived from SED modelling (see section~\ref{sec:SED_modelling}). This Figure also shows lines of constant temperature, which we take to define the boundaries between spectral types FGKM: $6000~\mathrm{K}$, $5200~\mathrm{K}$ and $3700~\mathrm{K}$. 

The control sample's characteristic distribution in Fig.~\ref{fig:sample_properties} results from the fact that the DEBRIS sample is volume-limited for each spectral type, and earlier-type stars, being less common, tend to be at greater distances. It is clear that the distance distributions of the two samples are very different, with the planet hosts tending to be considerably further away (since only some fraction of stars will be chosen as suitable targets for RV observations, and only some fraction of those targeted will have planets detected). This is not ideal, as the typical sensitivity of far infrared observations to emission from circumstellar dust will be considerably lower for the planet hosts. This difference in sensitivity means that the populations of debris discs in the two samples cannot simply be compared directly in terms of their relative detection rates. However, we can compare them using survival analysis (e.g. \citealt{Feigelson85_UpperLimits}), a set of methods which take account of upper limits on the fractional luminosities in the case of non-detections (see section~\ref{sec:results}).

In addition, it is apparent from Fig.~\ref{fig:sample_properties} that while the control sample is fairly uniformly distributed across the spectral types, the planet host sample is strongly biased towards G stars. Because of the well-established dependence of disc detection rate on spectral type (as discussed in section~\ref{sec:planet_sample}), the difference in effective temperature distributions must be considered when comparing the disc statistics of the two samples. In particular, a direct comparison of the full samples will make the disc fraction of the control sample appear too low relative to that of the planet sample, since the control sample has a greater proportion of late-type stars (see section~\ref{sec:results}).

\begin{figure}
	\centering
    \hspace{-0.5cm}
	\includegraphics[width=0.5\textwidth]{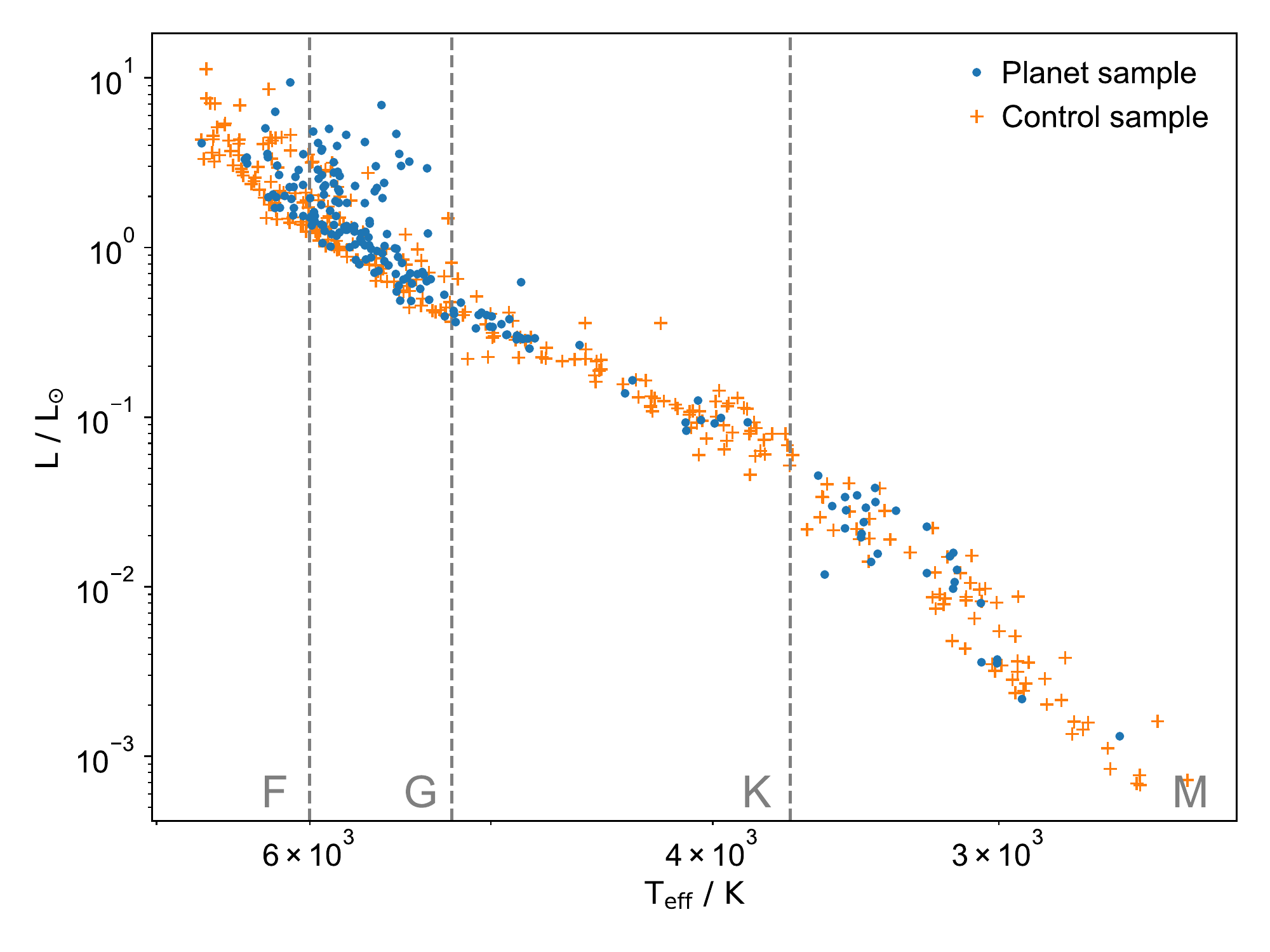}
	\caption{H-R diagram showing both our planet host sample and control sample. The stellar luminosities and effective temperatures are derived from SED modelling (see section~\ref{sec:SED_modelling}). The dashed lines show approximate boundaries between spectral types.}
	\label{fig:hr_diagram}
\end{figure}

Fig.~\ref{fig:hr_diagram} provides another comparison of the two samples, in the form of a H-R diagram. Both samples lie along the main sequence by construction, but the planet hosts tend to lie slightly further away from the zero age main sequence curve, suggesting that they are somewhat older on average. This could bias the planet host sample towards a lower rate of disc detection, and is likely a result of the fact that RV observations will preferentially target less active -- and therefore generally older -- stars. However, given that stellar ages are generally poorly constrained, we do not apply an age cut to our samples. We also see from Fig.~\ref{fig:hr_diagram} that both samples show a small number of systems with anomalously high luminosities. Such systems are close binaries, which also explains why they tend to be a factor of $\sim$2 brighter than expected for a single star, since in cases where the secondary star contributes non-negligible flux it must be of a similar spectral type to the primary. We note that close binaries will later be excluded from our analysis (see below).

\begin{figure}
	\centering
    \hspace{-0.5cm}
	\includegraphics[width=0.5\textwidth]{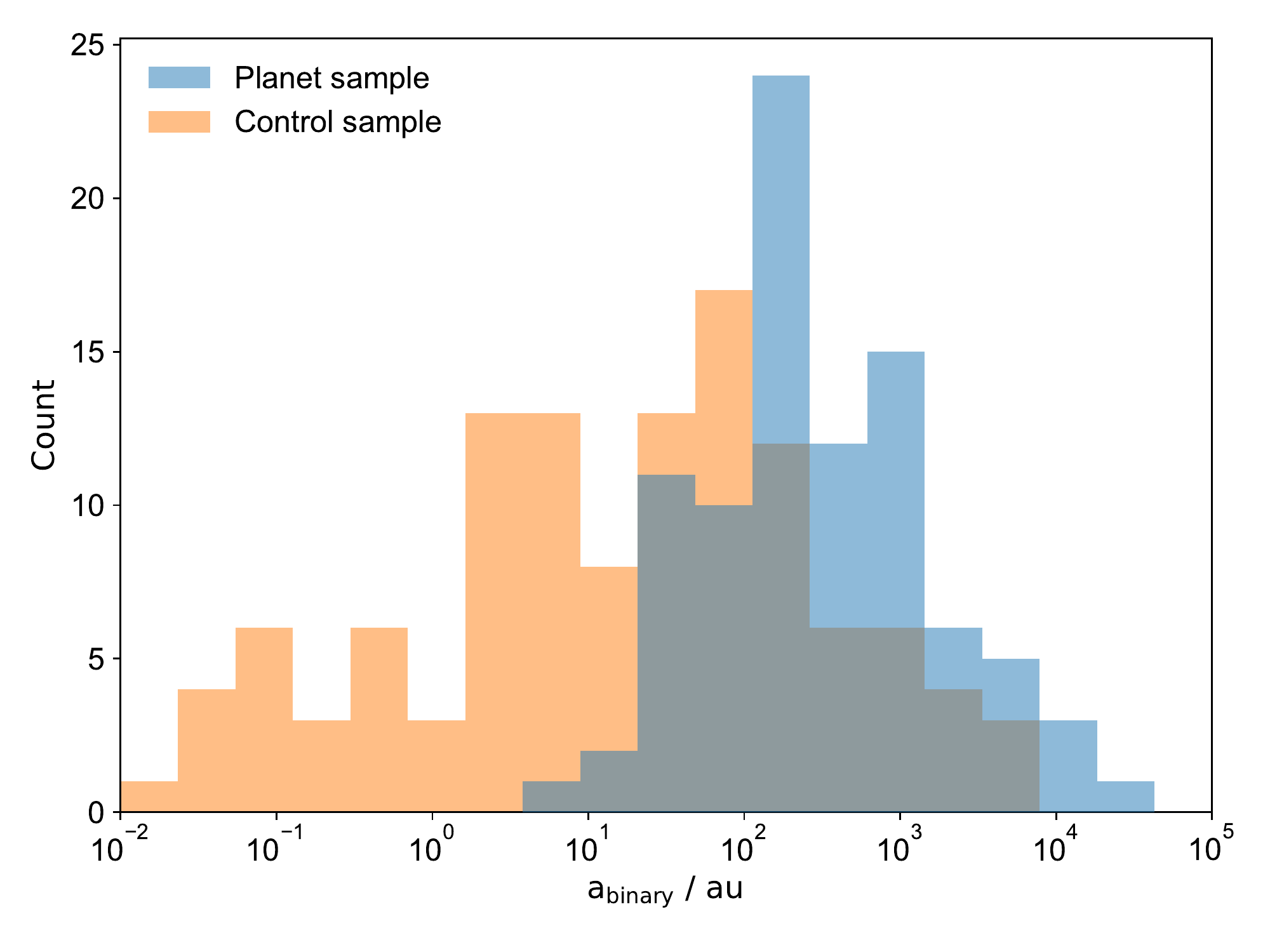}
	\caption{Distributions of binary separation for both our planet host sample and control sample. The control sample contains a much higher proportion of close binaries than the planet host sample.}
	\label{fig:binary_distributions}
\end{figure}

Finally, we compare the stellar separations of the binaries in the two samples in Fig.~\ref{fig:binary_distributions}. We see that the binaries in the planet host sample are on average much more widely separated than those in the control sample. This is not unexpected, as RV surveys will tend to avoid systems with close stellar companions, since such companions will induce their own strong RV signals, making the detection of planets difficult. The difference in binary separation distributions is another source of bias, driving the control sample towards a lower detection rate (see references in section~\ref{sec:planet_sample}), and so for a fair comparison close binaries should be excluded (see section~\ref{sec:results}).

With both stellar samples assembled, we proceed in the following section to explain how we identify and characterise their debris discs.

\section{SED Modelling}
\label{sec:SED_modelling}

This section outlines how we use the SEDs of the stars in our two samples to determine which systems we consider to host a debris disc, and for such systems, to estimate the dust temperature and fractional luminosity. We follow the procedure described by \citet{Yelverton19_BinaryDebris}, and refer the reader to that paper for a more detailed discussion of our method.

To identify discs, we search for excess infrared emission beyond what would be expected from the stellar photosphere, interpreting such emission as originating from dust in thermal equilibrium with the stellar radiation. We focus in particular on Kuiper belt analogues, with typical radii of tens of au, implying typical dust temperatures of $\sim$50~K (from equation~3 of \citealt{Wyatt08_Review}). Such discs are therefore best searched for at $\sim$100~$\mu$m (from equation~1 of \citealt{Wyatt08_Review}), which is why we constructed our samples to contain only systems which have been observed at 70 or 100~$\mu$m by MIPS and/or PACS. 

\begin{table*}
\begin{adjustbox}{max width=\textwidth,center}
\begin{tabular}{lccccccccccccccc}
\hline
  &&\multicolumn{4}{c}{MIPS 70~$\mu$m}  && \multicolumn{4}{c}{PACS 70~$\mu$m} && \multicolumn{4}{c}{PACS 100~$\mu$m} \\
 
 \cline{3-6}  \cline{8-11}  \cline{13-16} \\

\multirow{2}{*}{Name} & \multirow{2}{*}{Sample} & $F_{\nu,\mathrm{obs}}$ & $\sigma_{\mathrm{obs}}$ & $F_{\nu,\mathrm{pred}}$ & \multirow{2}{*}{$\chi$} && $F_{\nu,\mathrm{obs}}$ & $\sigma_{\mathrm{obs}}$ & $F_{\nu,\mathrm{pred}}$ & \multirow{2}{*}{$\chi$} && $F_{\nu,\mathrm{obs}}$ & $\sigma_{\mathrm{obs}}$ & $F_{\nu,\mathrm{pred}}$ & \multirow{2}{*}{$\chi$} \\  

&& / mJy & / mJy & / mJy &  && / mJy & / mJy & / mJy &  && / mJy & / mJy & / mJy &  \\ \hline

HD 142  & P &        &      & 12.96 &       && 16.05 & 0.85 & 13.68 & 2.78  &&       &      & 6.75  &       \\
HD 166  & C & 105.80 & 6.32 & 14.34 & 14.47 && 96.83 & 7.71 & 15.14 & 10.60 && 65.00 & 4.21 & 7.47  & 13.67 \\
HD 693  & C & 37.36  & 4.11 & 27.95 & 2.28  &&       &      & 29.50 &       && 18.87 & 1.84 & 14.46 & 2.39  \\
HD 739  & C & 20.11  & 2.78 & 17.30 & 1.01  &&       &      & 18.26 &       && 8.63  & 1.72 & 8.95  & -0.19 \\
HD 1237 & P & 11.81  & 2.08 & 9.32  & 1.19  &&       &      & 9.84  &       && 4.10  & 1.54 & 4.86  & -0.49 \\
HD 1326 & P & 30.50  & 2.64 & 30.78 & -0.10 &&       &      & 32.52 &       && 14.74 & 2.19 & 16.14 & -0.64 \\
HD 1461 & P & 60.39  & 6.79 & 8.63  & 7.63  &&       &      & 9.11  &       && 61.13 & 2.42 & 4.49  & 23.42 \\
HD 1581 & C & 69.96  & 6.72 & 59.15 & 1.60  &&       &      & 62.45 &       && 36.13 & 2.33 & 30.82 & 2.25  \\
HD 1835 & C & 20.46  & 5.33 & 8.81  & 2.18  &&       &      & 9.31  &       && 8.40  & 1.72 & 4.59  & 2.22  \\
HD 3443 & C & 19.23  & 5.12 & 22.44 & -0.62 &&       &      & 23.69 &       && 9.38  & 1.32 & 11.68 & -1.73 \\

\multicolumn{16}{c}{$\cdots$} \\
\hline
\end{tabular}\end{adjustbox}
\caption{Photometry in the MIPS 70~$\mu$m, PACS 70~$\mu$m and PACS 100~$\mu$m bands for all systems studied in this paper. In the second column, a P or C indicates that a star is in the planet host or control sample respectively. For each band, the columns show the observed flux density, its associated uncertainty, the predicted photospheric flux density, and the significance as defined in equation~(\ref{eqn:chidef}). Here we show only the first ten lines; a full machine-readable version can be obtained online.}
\label{tab:photometry}
\end{table*}

We compiled SEDs for all stars using archival photometry at wavelengths ranging from visible to far infrared. For the MIPS and PACS observations, we measured fluxes from the images using PSF fitting for unresolved sources and aperture photometry in cases where a disc is resolved. The resulting MIPS and PACS photometry is shown in Table~\ref{tab:photometry}. We then used the \textsc{sdf} code\footnote{Available at \url{https://github.com/drgmk/sdf}} to fit a model to each SED, consisting of a \textsc{PHOENIX} stellar component (\citealt{Allard12_PHOENIX}), as well as a modified black body component to fit any excess emission in cases where this improves the fit. The modified black body spectrum as a function of wavelength $\lambda$ is given by $\alpha B_{\nu}(\lambda,T)$ for $\lambda<\lambda_0$ and $\alpha (\lambda_0/\lambda)^{\beta}B_{\nu}(\lambda,T)$ for $\lambda>\lambda_0$, where $B_{\nu}$ is the Planck function, and the temperature $T$ and scaling factor $\alpha$, as well as the parameters $\lambda_0$ and $\beta$, are free parameters. We use this functional form to account for the fact that real circumstellar dust grains will not act as perfect black bodies, in particular emitting inefficiently at wavelengths longer than their own size; $\lambda_0$ and $\beta$ are thus related to the grain size distribution. In section~\ref{sec:results}, we will characterise the discs by their black body dust temperatures directly from the fit, as well as their fractional luminosities, derived by numerically integrating the dust spectrum and normalising by the integral of the stellar spectrum.

We allowed for two modified black body components where such a fit was favoured over a single-component model. The preference for two dust components is generally driven by the presence of a \textit{Spitzer} Infrared Spectrograph (IRS; \citealt{Houck04_IRS}) spectrum, which gives a strong constraint on the SED at wavelengths between $5$ and $37$~$\mu$m. Systems with a cool excess at 70 or 100~$\mu$m are thus sometimes better modelled with an additional warmer component that contributes to the SED over the IRS wavelength range, likely indicating that their discs have multiple spatial components (\citealt{Kennedy14_TwoTemp}). In such cases, we record the properties of the cooler component, since we are focusing on Kuiper belt analogues. Note that while two-temperature fits are sometimes favoured, we would obtain very similar results from single-temperature fits, since the additional warmer components do not contribute much flux to the SEDs at the wavelengths where the cooler components peak.

\begin{figure}
	\centering
    \hspace{-0.5cm}
	\includegraphics[width=0.5\textwidth]{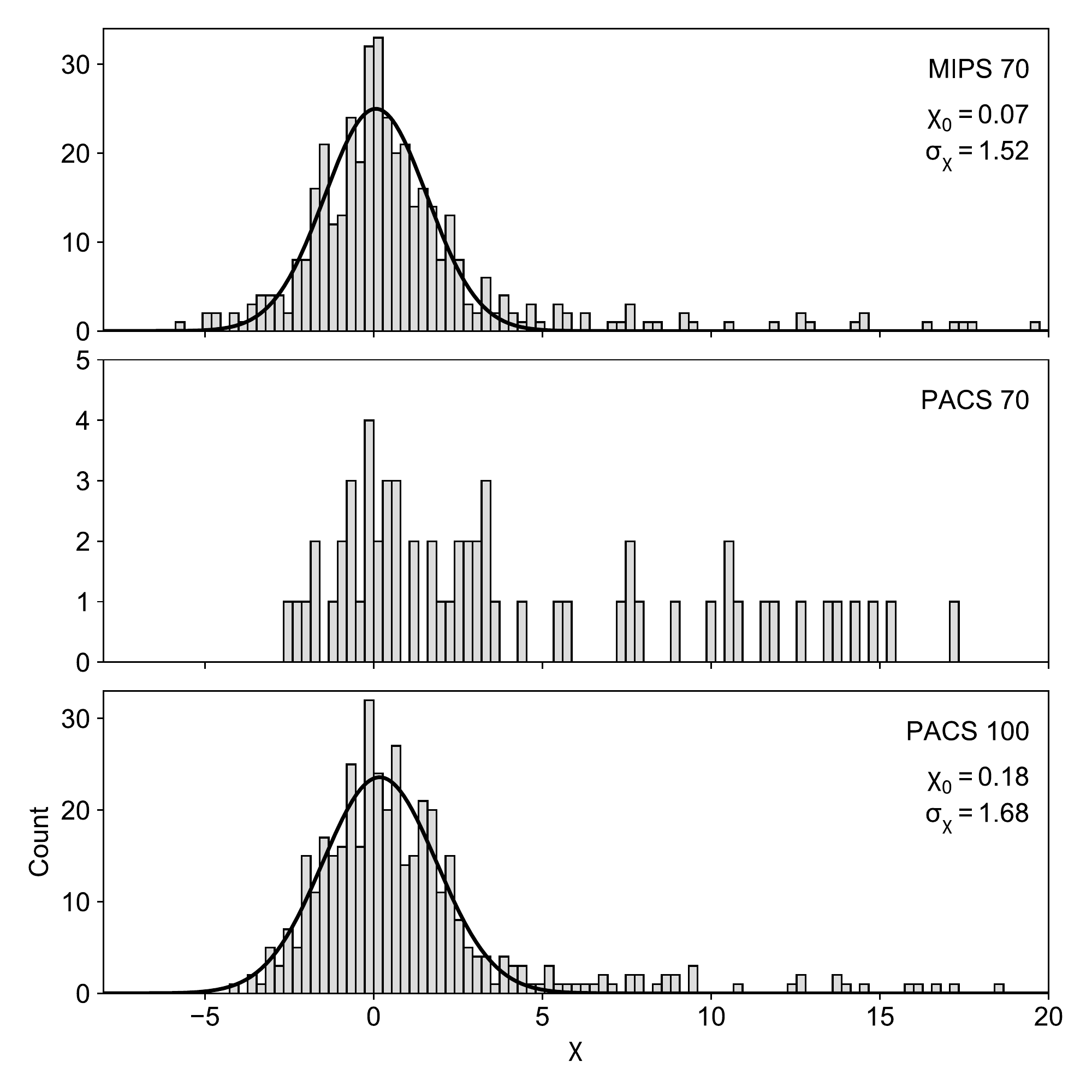}
	\caption{Distributions of photometric excess significance $\chi$ in the MIPS 70~$\mu$m and PACS 70 and 100~$\mu$m bands, for all systems studied in this paper (i.e. with the planet host and control samples merged). The bars show the empirical significances, while the curves in the upper and lower panels show best-fitting Gaussians with mean $\chi_0$ and standard deviation $\sigma_{\chi}$. We do not attempt to fit a Gaussian to the PACS 70~$\mu$m distribution, and instead assume that the underlying PACS 70 and 100~$\mu$m distributions are the same. Systems with $\chi>20$ are not shown.} 
	\label{fig:chi_histograms}
\end{figure}

Not all systems whose SED modelling favours the presence of a modified black body component are considered to have a \textit{significant} infrared excess. We define a criterion for an excess to be considered significant -- and therefore for a system to be considered a debris disc host -- based on the quantity $\chi$, defined as:

\begin{equation}\label{eqn:chidef}
    \chi=\frac{F_{\nu,\mathrm{obs}}-F_{\nu,\mathrm{pred}}}{\sqrt{\sigma_{\mathrm{obs}}^2+\sigma_{\mathrm{pred}}^2}},
\end{equation}

where $F_{\nu,\mathrm{obs}}$ is the observed flux density at a particular wavelength, $F_{\nu,\mathrm{pred}}$ is the flux density expected from the star alone at that wavelength (from the modelling described above), and $\sigma_{\mathrm{obs}}$ and $\sigma_{\mathrm{pred}}$ are the uncertainties on these quantities. Thus, the significance $\chi$ is the number of standard deviations by which the measured flux exceeds the photospheric flux at a given wavelength. We show the far infrared values of $F_{\nu,\mathrm{pred}}$ and $\chi$ along with the observed fluxes in Table~\ref{tab:photometry}; note that $\sigma_{\mathrm{pred}}$ is typically around 1\% of $F_{\nu,\mathrm{pred}}$. Fig.~\ref{fig:chi_histograms} shows the distributions of $\chi$ in the MIPS 70~$\mu$m and PACS 70 and 100~$\mu$m bands, for all stars considered in this paper. We choose to merge the planet host and control samples for the purpose of analysing the significances, since this increases the number of stars per histogram, and making separate histograms for each sample gives similar results (which is unsurprising; there is no reason to expect the samples' $\chi$ distributions to differ). Rather than simply defining a significant excess as one with $\chi>3$, we fit Gaussians to these distributions (also shown in Fig.~\ref{fig:chi_histograms}), with mean $\chi_0$ and standard deviation $\sigma_{\chi}$, then require $\chi>\chi_0+3\sigma_{\chi}$ in at least one of these bands. Using this criterion means that any inaccurately estimated uncertainties are accounted for. The distributions do in fact have standard deviations greater than unity, indicating that the uncertainties ($\sigma_{\mathrm{obs}}$ and/or $\sigma_{\mathrm{pred}}$) have been underestimated. This underestimation could for example be related to the measurements and/or predictions themselves, differences in photometric systems, or incorrect bandpasses being used to generate our synthetic photometry.

It is clear from Fig.~\ref{fig:chi_histograms} that a relatively small number of stars in our sample have been observed by PACS at 70~$\mu$m. In addition, the PACS 70~$\mu$m significance distribution is skewed towards positive $\chi$, a result of the fact that stars in the DEBRIS sample (from which our control sample is derived) were observed by default at 100~$\mu$m, with only potentially interesting sources later followed up at 70~$\mu$m (e.g. \citealt{Lestrade12_GJ581}). For these reasons, rather than fitting a Gaussian to the PACS 70~$\mu$m significances, we assume that the underlying distribution is the same as the PACS 100~$\mu$m distribution.

Having established a criterion to define a significant infrared excess and a method for deriving the disc temperature and fractional luminosity where such an excess is present, we show the results that follow in the next section.

\section{Results and Discussion}
\label{sec:results}

We begin this section by presenting the properties of the discs we identify in our two samples, then proceed to compare the samples statistically.

\subsection{Disc properties}
\label{sec:discproperties}

Using the criterion defined in section~\ref{sec:SED_modelling} based on the distributions in Fig.~\ref{fig:chi_histograms}, we detect circumstellar dust at a significant level around $26/201=13^{+3}_{-2}\%$ of the planet hosts and $25/294=9^{+2}_{-1}\%$ of the control stars, where the uncertainties quoted are binomial (e.g. \citealt{Burgasser03_BinomialErrors}). Recall from section~\ref{sec:sample_properties} that these detection fractions should not be compared directly, since they will be influenced by the distances, spectral types and binary separations of the stars, which differ considerably between the two samples. However, the fact that the detection rate is somewhat higher for the planet hosts despite their greater distances hints at a potentially interesting result.

\begin{table}

\begin{adjustbox}{max width=0.5\textwidth,center}
\begin{tabular}{lcrrrr}
\hline
Name & Sample & $T$ / K  & $\sigma_{T}$ / K & $f$ & $\sigma_{f}$ \\
\hline

HD 166   & C & 79  & 2  & 4.9E-05 & 1.4E-06 \\
HD 1461  & P & 57  & 1  & 4.3E-05 & 1.5E-06 \\
HD 5133  & C & 62  & 4  & 1.1E-05 & 1.1E-06 \\
HD 10647 & P & 40  & 1  & 2.4E-04 & 8.8E-06 \\
HD 10700 & P & 104 & 9  & 1.5E-05 & 2.8E-06 \\
HD 16673 & C & 94  & 11 & 5.9E-06 & 5.5E-07 \\
HD 17925 & C & 81  & 10 & 3.5E-05 & 6.1E-06 \\
HD 22049 & P & 51  & 11 & 4.2E-05 & 2.2E-05 \\
HD 22484 & C & 88  & 6  & 9.4E-06 & 7.7E-07 \\
HD 23356 & C & 61  & 3  & 1.7E-05 & 1.4E-06 \\

\multicolumn{6}{c}{$\cdots$} \\
\hline
\end{tabular}\end{adjustbox}
\caption{List of all systems we find to have a significant infrared excess, along with the best fitting black body temperatures and fractional luminosities of their discs (and their uncertainties), derived from the SED modelling outlined in section~\ref{sec:SED_modelling}. In the second column, a P or C indicates that a star is in the planet host or control sample respectively. Here we show only the first ten lines; a full machine-readable version can be obtained online.}

\label{tab:discparams}
\end{table}

\begin{figure}
	\centering
    \hspace{-0.5cm}
	\includegraphics[width=0.5\textwidth]{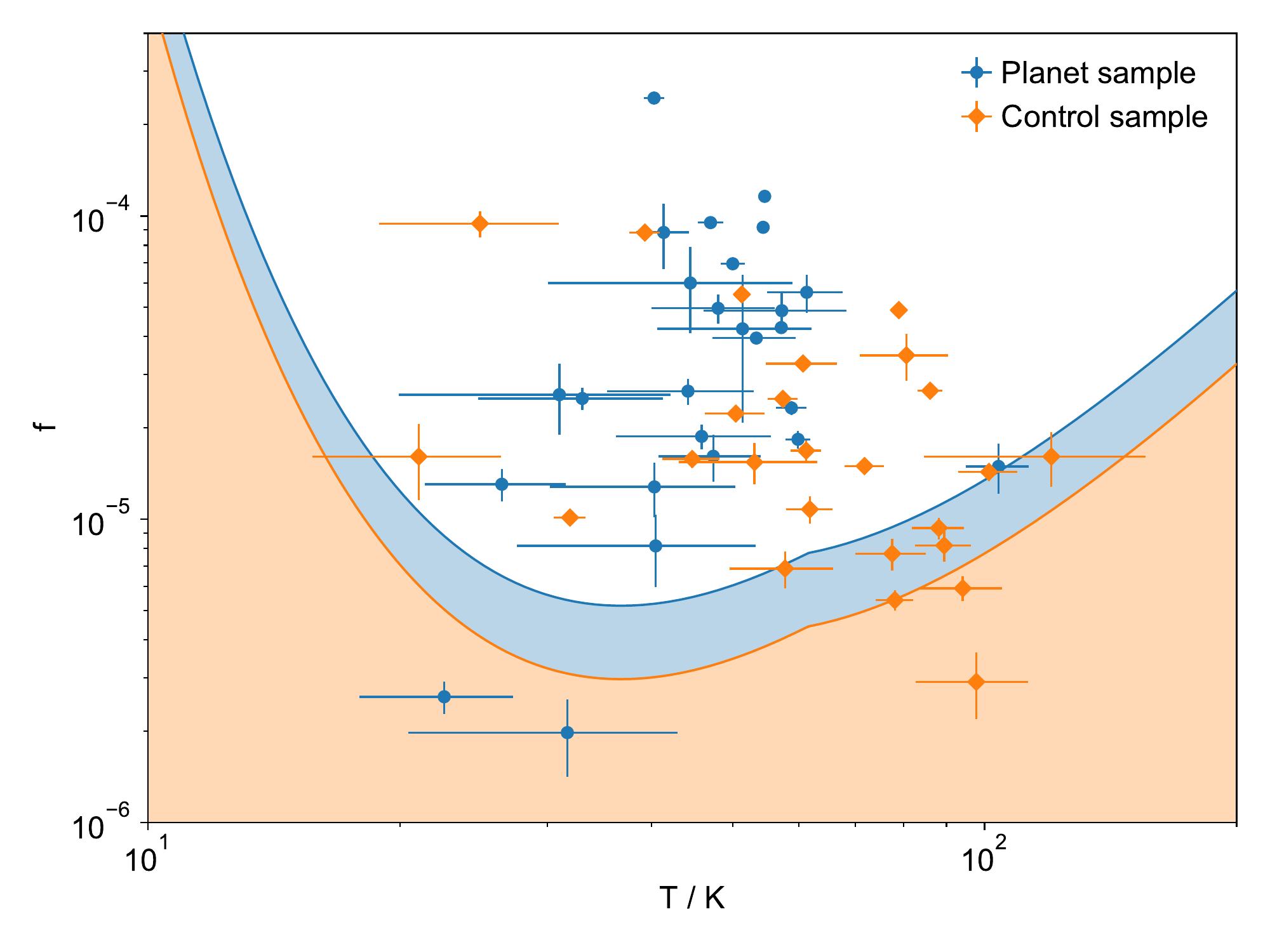}
	\caption{Fractional luminosity against dust temperature for the discs detected in our two samples. For two-temperature systems, we show only the cooler component (see section~\ref{sec:SED_modelling}).  The shaded regions show where a disc would be undetectable around a star with the median luminosity and at the median distance for each sample.}
	\label{fig:fvsT}
\end{figure}

The fractional luminosities $f$ and dust temperatures $T$ of the discs we detect are listed in Table~\ref{tab:discparams} and plotted in Fig.~\ref{fig:fvsT}. This Figure also shows illustrative detection thresholds for both samples. These were calculated using equation~(8) of \citet{Wyatt08_Review}; we took the sensitivity to be 5~mJy, which is typical for PACS\footnote{See Table~3.2 of the PACS Observer's Manual, \url{http://herschel.esac.esa.int/Docs/PACS/html/pacs_om.html}}, and the thresholds shown for each sample correspond to a star with that sample's median luminosity $L$ and at its median distance $d$. These medians are 1.0~$L_\odot$ and 29~pc for the planet sample, and 0.4~$L_\odot$ and 14~pc for the control sample. Note that $d$ and $L$ appear in the equation for the minimum detectable $f$ in the combination $d^2L^{-1}$. The fact that the lines plotted in Fig.~\ref{fig:fvsT} are spaced only by a factor of $\sim$2 is a result of the fact that the control stars are closer, but also of a later spectral type (i.e. lower luminosity) on average. Comparing only stars of the same spectral type, their spacing would be a factor of $\sim$4. The kinks in the lines at $\sim$60~K result from the fact that we calculate the thresholds at both 70 and 100~$\mu$m and plot the lower of the two. Detections may (and do) in fact lie below the lines, since some stars will be sufficiently nearby and/or luminous that their thresholds are lower than those shown.

Note that for systems with a significant excess in only one far infrared band, the $f$ and $T$ values are still constrained by non-detections in other bands, which explains why the error bars in Fig.~\ref{fig:fvsT} are of a reasonable size. As an example, HD~72659 (with $T=40~\mathrm{K}$ and $f=1.3\times10^{-5}$) has a $\chi=8.6$ detection in PACS 100~$\mu$m, but a low MIPS 70~$\mu$m significance of $\chi=1.6$ (and no PACS 70$~\mu$m data). Nonetheless, the model is constrained on the warm side by the MIPS 70~$\mu$m photometry, and on the cool side by PACS 160~$\mu$m (which has $\chi=1.2)$; consequently, its $f$ and $T$ values are constrained to within $\sim$25\%. The majority of systems with significant excesses in our samples have photometry at two or more far infrared wavelengths (including 160~$\mu$m). The two systems which do not are HD~104067 and HD~150706; neither of these has been observed by PACS at any wavelength, but both have IRS spectra which, combined with their MIPS 70~$\mu$m photometry, allow their disc parameters to be well constrained.

\subsection{Fractional luminosity comparison}
\label{sec:fraclumcomparison}

The aim of this paper is to determine whether the debris disc properties of planet hosts are statistically distinguishable from those of stars without known planets, and we now turn our attention to this question. We first assess whether the fractional luminosity distributions of the two populations differ; in particular, whether it is possible to show that planet hosts tend to have brighter discs, as has been predicted theoretically (e.g. \citealt{Wyatt07_Metallicity}; \citeauthor{Raymond11_TerrestrialPlts1} \citeyear{Raymond11_TerrestrialPlts1}, \citeyear{Raymond12_TerrestrialPlts2}) and hinted at by previous observational studies (e.g. \citealt{Bryden09_PlanetDisc}; \citealt{Wyatt12_LowMassPlts}; \citealt{Marshall14_Correlations}; \citealt{Matthews14_DiscReview}; \citealt{MoroMartin15_PlanetsAndKBs}). As discussed in section~\ref{sec:sample_properties}, since the distance distributions (and hence sensitivities to dust) of the two samples are different, rather than directly comparing the fractional luminosity distributions of the detected discs, we must use survival analysis (e.g. \citealt{Feigelson85_UpperLimits}), which can estimate the underlying distributions in a way that takes into account the fractional luminosity upper limits of the non-detections.

To estimate these upper limits, for each system without a significant infrared excess we first calculate the minimum ratio of observed to predicted stellar flux that would have resulted in a detection in each of the far infrared bands (i.e. MIPS 70~$\mu$m, PACS 70~$\mu$m and PACS 100~$\mu$m) in which that system has an observation. From equation~(\ref{eqn:chidef}), combined with our excess criterion $\chi>\chi_0+3\sigma_{\chi}$, in any particular band this ratio is given by 

\begin{equation}\label{eqn:fluxratiothreshold}
    \left. \frac{F_{\nu,\mathrm{obs}}}{F_{\nu,\mathrm{pred}}}\right|_{\mathrm{lim}} = \frac{(\chi_0+3\sigma_{\chi}) {\sqrt{\sigma_{\mathrm{obs}}^2+\sigma_{\mathrm{pred}}^2}}}{F_{\nu,\mathrm{pred}}} + 1.
\end{equation}

We then use equation~(11) of \citet{Wyatt08_Review} to convert this flux ratio detection threshold into a minimum detectable fractional luminosity for each band. The fractional luminosity threshold $f_{\mathrm{lim}}$ in fact depends on dust temperature $T$, which has no constraint when no disc is detected. For the upper limits we choose to take the value of $f_{\mathrm{lim}}$ at the minimum of the $f_{\mathrm{lim}}(T)$ curve (i.e. at temperatures of $\sim$35--50~K); these are recorded in Table~\ref{tab:upperlimits}. Thus, any disc with a fractional luminosity below our nominal upper limit is guaranteed to be undetected, but it is also possible for a disc with a somewhat higher fractional luminosity to remain undetected if its temperature is sufficiently cold or hot. We discuss the effect of alternative dust temperature assumptions at the end of section~\ref{sec:allplanets}. For systems with observations in more than one of the far infrared bands, we record the smallest of the calculated per-band upper limits. While comparing fractional luminosities rather than flux ratios directly has the disadvantage that it is impossible to calculate true upper limits due to the temperature dependence, we prefer to do this since it provides a way to place the different photometric bands on a more equal footing.

\begin{table}
\centering
\begin{tabular}{lcr}
\hline
Name & Sample & $f_{\mathrm{lim}}$\\
\hline

HD 142  & P & 2.40E-06 \\
HD 693  & C & 1.70E-06 \\
HD 739  & C & 2.30E-06 \\
HD 1237 & P & 6.30E-06 \\
HD 1326 & P & 1.10E-05 \\
HD 1581 & C & 1.20E-06 \\
HD 1835 & C & 6.10E-06 \\
HD 3443 & C & 2.30E-06 \\
HD 3651 & P & 2.70E-06 \\
HD 4203 & P & 1.00E-04 \\

\multicolumn{3}{c}{$\cdots$} \\
\hline
\end{tabular}
\caption{Upper limits on the fractional luminosities of discs in systems without a significant infrared excess. See section~\ref{sec:fraclumcomparison} for details of how these are calculated. In the second column, a P or C indicates that a star is in the planet host or control sample respectively. Only the first ten lines of the table are reproduced here; a full machine-readable version is available online.}
\label{tab:upperlimits}
\end{table}

\subsubsection{Comparison irrespective of planet mass}
\label{sec:allplanets}

\begin{figure}
	\centering
    \hspace{-0.5cm}
	\includegraphics[width=0.5\textwidth]{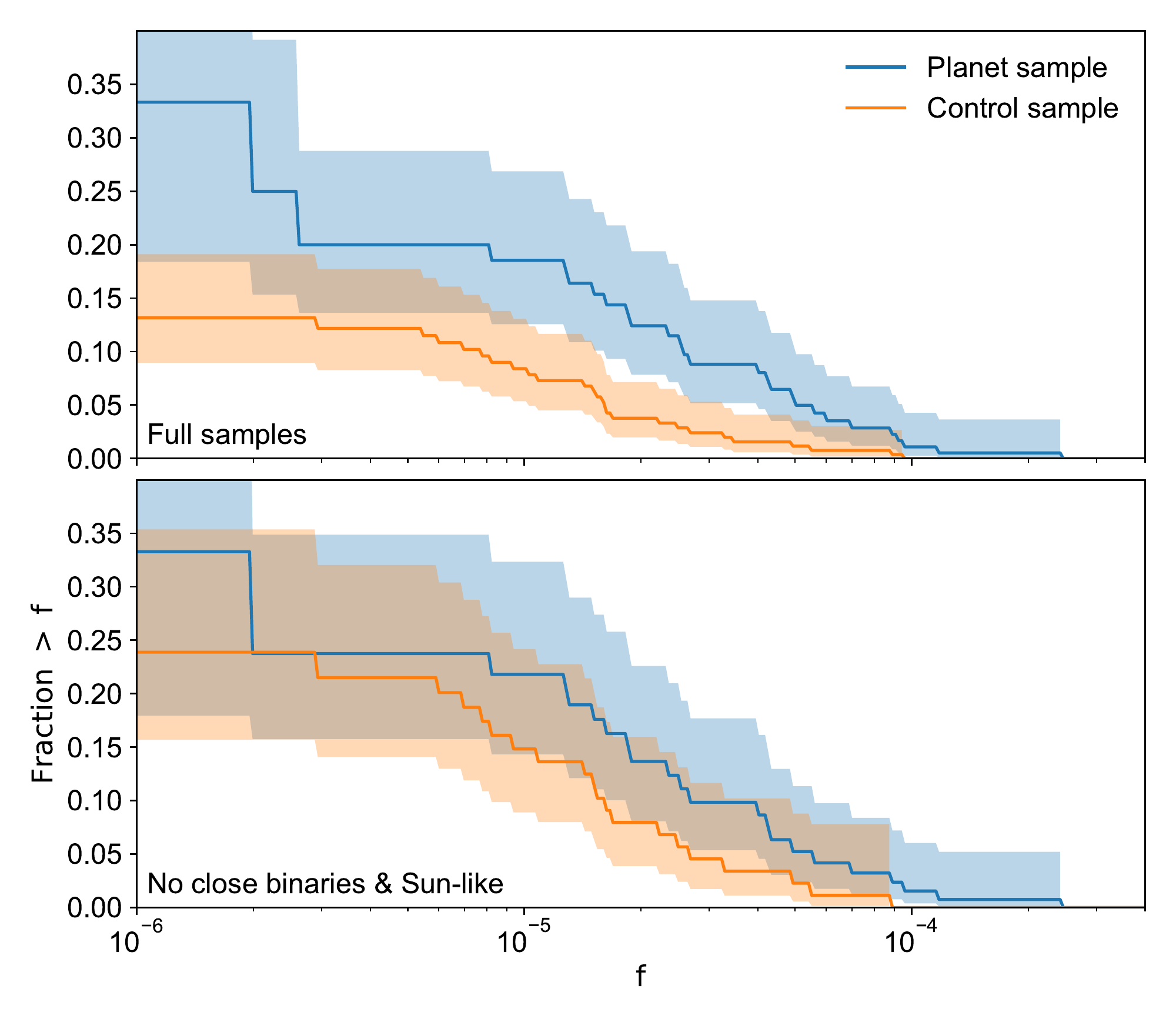}
	\caption{Cumulative distribution functions of fractional luminosity for both our planet host and control samples, calculated using the Kaplan-Meier product-limit estimator. The value on the vertical axis at a fractional luminosity $f$ shows the fraction of stars with a detected debris disc brighter than $f$; the distributions do not reach unity because not all systems have a disc detected. Note that the distributions are constant over ranges of $f$ where no discs are detected. The shaded regions indicate the 95\% confidence intervals. The upper panel shows a comparison of the full samples, while in the lower panel we have excluded binaries closer than 135~au and retained only stars with effective temperatures between 4700 and 6300~K. The planet host stars tend to have brighter discs than those without known planets, but this difference is not significant when the cuts on binarity and spectral type described above are made.}
	\label{fig:KM_estimators}
\end{figure}

We use the Kaplan-Meier product-limit estimator (\citealt{Kaplan58_KMEstimator}), as implemented in the \textsc{lifelines} package by \citet{Lifelines}, to estimate the fractional luminosity cumulative distribution function (CDF) for each sample, using both the detections and upper limits. These are shown in the upper panel of Fig.~\ref{fig:KM_estimators}. We note that this panel can be compared with Fig.~4 of \citet{Matthews14_DiscReview}. Although their plot shows 100~$\mu$m flux ratios rather than fractional luminosities, and has been generated in a way that does not guarantee monotonic CDFs (in contrast to the Kaplan-Meier estimator), our results are qualitatively similar. The fact that the planet host CDF lies above the control CDF everywhere indicates that stars with planets do tend to host brighter debris discs than those without known planets. We can assess the significance of this result using the five different statistical tests included in the \textsc{asurv} package (\citealt{Lavalley92_ASURV}); full details of the tests can be found in \citet{Feigelson85_UpperLimits}. For this work, the Peto \& Prentice test is most appropriate, since our samples have different sizes and typical upper limits (\citealt{Latta81_TwoSampleTests}). Each test returns a $p$-value indicating the probability of obtaining empirical distributions which differ at least as much as those obtained, under the hypothesis that the true underlying distributions are the same. Thus, a small $p$-value would suggest that the fractional luminosities of the two samples differ significantly. The $p$-values from the five different tests are shown in the top row of Table~\ref{tab:pvalues}. All tests give similar results, with all $p$-values below $\sim$0.003, indicating that the full samples have fractional luminosity distributions that differ with greater than 3$\sigma$ confidence, as stated in \citet{Matthews14_DiscReview}. However, this difference cannot necessarily be attributed to the presence of planets, since the two samples have other inherent differences which are known to impact debris disc properties, as discussed in section~\ref{sec:sample_properties}.

\begin{table*}

\begin{adjustbox}{max width=\textwidth,center}
\begin{tabular}{rccccccc}

 & & & \multicolumn{5}{c}{$p$-values} \\
 \cline{4-8}  \\
 
\multirow{2}{*}{Samples}&\multirow{2}{*}{$N_{\mathrm{p}}$}  &\multirow{2}{*}{$N_{\mathrm{c}}$} & Gehan  & Gehan  & \multirow{2}{*}{Logrank} & \multirow{2}{*}{Peto \& Peto} & \multirow{2}{*}{Peto \& Prentice} \\
& & & permutation variance & hypergeometric variance & & &  \\
\hline

Full    & 201 & 294 &  0.0018 & 0.0012 & 0.0022 & 0.0032 & 0.0019 \\
NCB     & 174 & 205 & 0.0092 & 0.0077 & 0.0246 & 0.0222 & 0.0181 \\
Sun-like         & 158 & 140 & 0.0794 & 0.0730 & 0.0704 & 0.0735 & 0.0658\\
Sun-like \& NCB  & 138 & 88 & 0.2501 & 0.2586 & 0.3696 & 0.3128 & 0.3206 \\       
\hline
\end{tabular}\end{adjustbox}
\caption{Summary of the statistical tests we applied to the fractional luminosities of the samples and subsamples. Each row corresponds to a different pair of subsamples. The subsamples are indicated in the first column; here NCB stands for no close binaries. See the text for more details on subsample selection. $N_{\mathrm{p}}$ and $N_{\mathrm{c}}$ give the number of stars in the relevant planet host and control subsamples respectively. The remaining columns show the $p$-values from each of the statistical tests.}
\label{tab:pvalues}
\end{table*}

To quantify the effects of the differences between the samples, we next consider various subsamples. Firstly, we cut out close binaries from both samples, as it is known that debris discs are less commonly detectable in such systems (see references in section~\ref{sec:planet_sample}). We define a close binary as one with a separation below 135~au, based on the results of \citet{Yelverton19_BinaryDebris}. The $p$-values (and sample sizes) for the resulting subsamples are shown in the second row of Table~\ref{tab:pvalues}; they are around an order of magnitude higher than for the full samples, reducing the confidence that the fractional luminosities are different to $\sim$2$\sigma$. This change in significance can be understood from Fig.~\ref{fig:binary_distributions}: since close binaries make up a higher proportion of the control stars than the planet hosts, cutting them out of both samples will increase the detection rate within the control sample more than for the planet host sample. That is, the CDF curves in the upper panel of Fig.~\ref{fig:KM_estimators} will both be scaled to higher values, but the gap between them will become smaller; the 95\% confidence intervals will also become larger due to the reduced sample sizes following the cuts.

Next, we consider subsamples of only Sun-like stars. Based on Fig.~\ref{fig:sample_properties}, we define a Sun-like star as one with an effective temperature between 4700 and 6300~K, since this is where the temperature distributions of the two samples are most similar, allowing the comparison between the samples to be as fair as possible in terms of spectral type. The results of the statistical tests on these subsamples are shown in the third row of Table~\ref{tab:pvalues}; again, the confidence is reduced below the full-sample case, to just below the 2$\sigma$ level. This is to be expected: the control sample contains a higher proportion of later-type stars, which have a lower disc detection rate (see references in section~\ref{sec:planet_sample}), so restricting to Sun-like stars only will raise the CDF of the control sample more than that of the planet host sample (qualitatively similar to the effect of removing close binaries).

It is clear that both binarity and spectral type greatly influence our conclusion about the significance of our result, and so for the most reliable comparison we should make both of the cuts outlined above simultaneously. The resulting $p$-values (shown in the fourth row of Table~\ref{tab:pvalues}) are high, reducing the confidence that the fractional luminosities are different to barely 1$\sigma$. We also show the Kaplan-Meier estimates of the CDFs with the binarity and spectral type cuts applied in the lower panel of Fig.~\ref{fig:KM_estimators}. While the smaller sample sizes following the cuts will tend to make the $p$-values higher, it is clear from inspection of the CDFs in the upper and lower panels of Fig.~\ref{fig:KM_estimators} that the higher $p$-values are not \textit{only} a result of the smaller samples: the planet host CDF does not change much, but the control CDF is scaled up by a factor of $\sim$2 in the lower panel, bringing the two curves close together such that the CDFs overlap within the 95\% confidence intervals. Thus, we conclude that even when taking advantage of data from far-infrared surveys that specifically targeted planet hosts to construct a large sample thereof (compared with e.g. \citealt{MoroMartin15_PlanetsAndKBs}), there is no evidence that debris discs are significantly brighter around stars with known planets than those without, once the inherent differences in binarity and spectral type of the two populations are taken into account.

Note that it is very likely that some stars in the control sample host planets which have not been detected simply because the stars have not been targeted by RV instruments; this fact will dilute any planet-disc correlation which is present. Using data from the HARPS RV survey, \citet{Mayor2011_38858Planet} estimated that $\sim$60--80\% of Sun-like stars host detectable planets with periods shorter than 10 years, whereas only 22/110 = 20\% of Sun-like stars without a close binary companion in the DEBRIS sample from which our control sample is drawn have a planet listed in NEA. This is why it would be ideal to use a sample of stars with published RV non-detections, but as discussed in section~\ref{sec:control_sample}, such a sample has only been published relatively recently (\citealt{Howard16_PltNonDetection}), and therefore will not have been targeted by far infrared surveys.

Finally, we consider how our results would be affected by making a different assumption about the temperatures of undetected discs. The fractional luminosity upper limits in Table~\ref{tab:upperlimits} were calculated at the minimum of the detection threshold curves, which is equivalent to assuming dust temperatures of $\sim$35 and $\sim$50~K at observation wavelengths of 100 and 70~$\mu$m. An alternative approach would be to assume that the undetected dust temperatures are equal to the median temperatures of the discs in each sample. These medians are 47 and 62~K for the planet host and control samples respectively. Thus, the upper limits for the planet sample using this method would be very similar to those that we calculated. For the control sample, the higher median temperature means that the fractional luminosity limits would be somewhat greater than our tabulated values, but only by a small amount (from Fig.~\ref{fig:fvsT}, the 100~$\mu$m sensitivity curve rises by a factor of only $\sim$1.5 from 35 to 62~K). Thus, the effect of the proposed alternative assumption on the CDFs in Fig.~\ref{fig:fvsT} would be to move the control sample curve slightly up relative to the planet sample curve, which would further decrease the significance of the difference between the two distributions. Thus, our conclusion that there is no significant difference between the two fractional luminosity distributions would not change. In any case, we favour the approach of taking the minimum of the sensitivity curve because it is not known whether undetected discs have the same median temperature as those which are detected.

\subsubsection{Effect of planet mass}
\label{sec:planetmass}

Here we investigate whether there is any evidence for a difference in the disc detection statistics between stars with low-mass and high-mass planetary systems. Following \citet{Wyatt12_LowMassPlts}, we divide our planet host sample into two subsamples distinguished by the presence or absence of a planet with $M\sin(i)$ greater than a Saturn mass (i.e. 0.3~$M_{\mathrm{J}}$), since theoretical arguments have been made for giant planets providing unfavourable conditions for the long-term survival of debris discs (\citeauthor{Raymond11_TerrestrialPlts1} \citeyear{Raymond11_TerrestrialPlts1}, \citeyear{Raymond12_TerrestrialPlts2}; \citealt{Wyatt12_LowMassPlts}).

\begin{figure}
	\centering
    \hspace{-0.5cm}
	\includegraphics[width=0.5\textwidth]{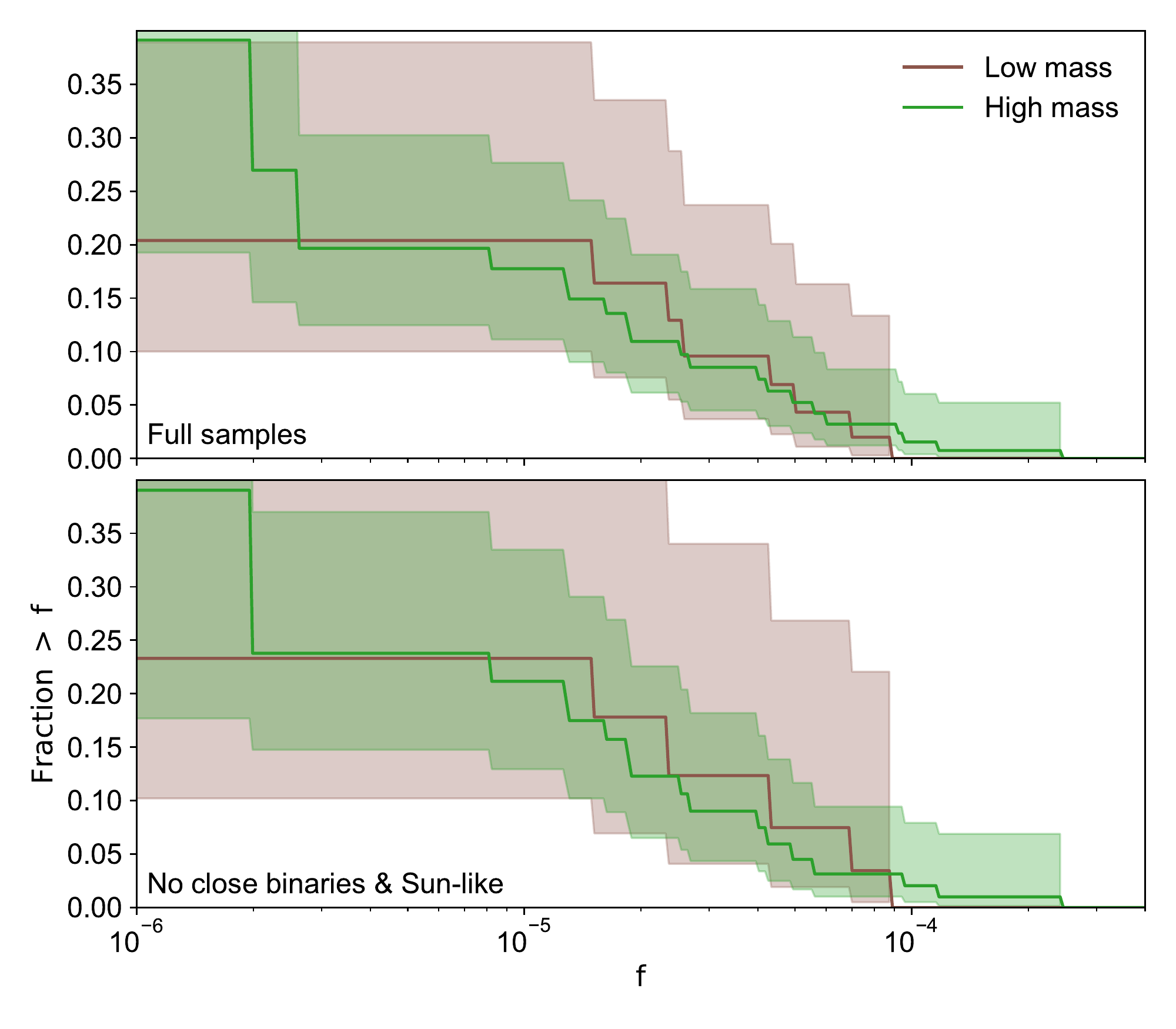}
	\caption{Same as Fig.~\ref{fig:KM_estimators}, but comparing the fractional luminosity distributions of subsamples of the planet host sample for which the maximum $M\sin(i)$ is either below or above the mass of Saturn.}
	\label{fig:KM_estimators_mass}
\end{figure}

For the low-mass planet hosts, we detect a disc around $7/62=12^{+5}_{-3}\%$ stars, while for the high-mass systems the detection rate is $19/139=14^{+3}_{-2}\%$. If we exclude close binaries and consider only Sun-like stars, these numbers become $5/34=15^{+8}_{-4}\%$ and $16/104=15^{+4}_{-3}\%$ for the low-mass and high-mass systems respectively. These fractions are all consistent within their uncertainties, and we thus conclude that detectable Kuiper belt analogues are in fact equally common in systems with and without giant planets detected via RV. Note that a direct comparison of detection rates is reasonable here, since the low-mass and high-mass subsamples are both drawn from the same larger sample and will thus have similar typical sensitivities. Using survival analysis to compare the fractional luminosities of the low-mass and high-mass subsamples similarly suggests that the disc populations are indistinguishable. The Kaplan-Meier CDFs shown in Fig.~\ref{fig:KM_estimators_mass} are clearly very similar for both subsamples regardless of whether we make the binarity and spectral type cuts, and the \textsc{asurv} statistical tests return $p$-values greater than 0.9.

As a check on our results, we can compare them against \citet{Wyatt12_LowMassPlts}, who studied samples of six low-mass and five high-mass planetary systems, all of which were G-type stars. They found excesses around 4/6 of the low-mass and 0/5 of the high-mass systems. For the low-mass systems, we agree that HD~102365 and HD~136352 do not have discs. Of those that \citet{Wyatt12_LowMassPlts} did find to have discs, we arrive at the same conclusion for 61~Vir (HD~115617). HD~69830 is well known to harbour hot dust (\citealt{Lisse07_HD69830}; \citealt{Marshall14_Correlations}) and does have an excess at $\sim$20~$\mu$m, but not in any of the three far infrared bands we focus on. Thus, though the system does contain circumstellar dust, the dust is of a different (likely transient) nature to the Kuiper belt analogues we focus on, and we do not classify it as a debris disc host for the purposes of this paper. We find that HD~20794 has a $\chi$ value of 3.7 in MIPS 70~$\mu$m and 4.2 in PACS 100~$\mu$m; it is known that this system does in fact host a marginally resolved disc with $f\sim 10^{-6}$ (\citealt{Kennedy15_KBSuperEarth}), but based solely on the $\chi$ distributions in Fig.~\ref{fig:chi_histograms} we cannot classify it as having significant excess emission in a consistent way. Since our definition of excess emission is somewhat conservative, it is inevitable that there will be some previously identified faint discs that we do not detect. Finally, HD~38858 is not in our planet host sample, since it is not listed as having a planet in NEA. This may be because \citet{Flores18_HD38858} have demonstrated that its originally identified planet candidate (\citealt{Mayor2011_38858Planet}) is an artefact of periodic stellar activity; note that \citet{Kennedy15_KBSuperEarth} stated that RV data shows evidence for a different planet in the system, but the analysis that led to this conclusion has not been published. For the high-mass planet hosts from \citet{Wyatt12_LowMassPlts}, we agree that HD~147513, 47~UMa (HD~95128), mu~Ara (HD~160691) and 51~Peg (HD~217014) do not host discs. HD~190360 is not in our planet host sample since we specifically excluded it due to infrared cirrus contamination (see section~\ref{sec:planet_sample}).

To summarise, of the nine \citet{Wyatt12_LowMassPlts} planet hosts which are also in our sample, we agree with their conclusion on excess infrared emission in seven cases. For the remaining two stars, they identify discs which we do not; this can be explained in one case (HD~69830) by the unusually high temperature of the dust, and in the other (HD~20794) by the fact that the excess is relatively weak and not formally significant when compared with the other stars in our sample, as evidenced in Fig.~\ref{fig:chi_histograms}. Since hot dust is rare, and the issue of borderline excesses applies equally to both low-mass and high-mass planetary systems, the fact that we find comparable detection rates for the two subsamples, in contrast to \citet{Wyatt12_LowMassPlts}, is likely simply a result of our much larger sample sizes rather than a difference in our methods. Note, though, that while we do have larger samples, the number of disc detections (particularly for the low-mass planets) is still small.

\subsubsection{Contamination by low-inclination stellar companions?}
\label{sec:stellarcompanions}

In this subsection we consider how the fact that the masses of radial velocity planets are degenerate with their inclinations might affect our results. As discussed in section~\ref{sec:planet_sample}, this fact means that some of the planet candidates in our sample may actually be stars on low-inclination orbits (as is likely the case for HD~114762, from \citealt{Kiefer09_HD114762}). One way in which we can assess this issue is by obtaining an alternative mass estimate using astrometry. \citet{Kervella19_PMa} tabulated proper motion anomalies for most \textit{Hipparcos} (\citealt{vanLeeuwen07_Hipparcos}) stars, where the proper motion anomaly is defined as the proper motion from either \textit{Hipparcos} or $Gaia$ Data Release 2 (\citealt{Gaia18_DR2Catalog}) minus the long-term motion between the $Hipparcos$ and $Gaia$ epochs. These proper motion anomalies can be simply converted into tangential velocity anomalies $\Delta v_t$ using the stellar distances. In Table~\ref{tab:anomalymass}, we list all systems in our samples which appear in \citet{Kervella19_PMa} with a significant anomaly (i.e. with $\Delta v_t$ greater than three times its uncertainty), determined by cross-matching against their catalogue. We calculate $\Delta v_t$ twice for each system, using both \textit{Hipparcos} and \textit{Gaia} proper motions; in cases where both are significant, the value listed in Table~\ref{tab:anomalymass} corresponds to the anomaly that implies the higher companion mass (see below). Note that 8/201 of the planet hosts and 50/294 of the control stars do not appear in \citet{Kervella19_PMa}; such stars are either too faint to appear in the \textit{Hipparcos} catalogue, or brighter than the \textit{Gaia} saturation limit. The latter of these issues particularly affects the control stars, since they are nearby. Assuming that the anomalies are caused by orbiting companions, we then use equation~(15) of \citet{Kervella19_PMa} to calculate the minimum mass required to explain each star's anomaly. The required mass $M_{\mathrm{anom}}$ of the companion depends on its orbital radius $r$; we record the mass at the minimum of the $M_{\mathrm{anom}}(r)$ curve (see Fig.~14 of \citealt{Kervella19_PMa} for an example of such a curve), which is typically equivalent to assuming an orbital radius of a few au (note that $M_{\mathrm{anom}}$ tends to be relatively insensitive to $r$ in the range $\sim$1--10~au, where many of the planets in our sample lie). Thus, stars with companions which are very close or very wide will have companion masses much greater than those listed in Table~\ref{tab:anomalymass}.

\begin{table}

\begin{adjustbox}{max width=0.5\textwidth,center}
\begin{tabular}{lcrrc}
\hline
Name & Sample & $\Delta v_t$ / ms$^{-1}$ & $M_{\mathrm{anom}}$ / $M_{\mathrm{J}}$ & Known binary\\
\hline

HD 142   & P & 122.1 & 12.5  & Y \\
HD 693   & C & 574.1 & 58.9  & N \\
HD 739   & C & 708.5 & 130.2 & N \\
HD 1326  & P & 23.5  & 2.7   & Y \\
HD 1581  & C & 322.6 & 29.6  & N \\
HD 4676  & C & 56.4  & 16.6  & Y \\
HD 4747  & C & 301.2 & 48.1  & Y \\
HD 8673  & P & 98.3  & 10.6  & Y \\
HD 9826  & P & 93.1  & 6.2   & Y \\
HD 10307 & C & 774.9 & 172.8 & Y \\

\multicolumn{5}{c}{$\cdots$} \\
\hline
\end{tabular}\end{adjustbox}
\caption{Systems with a significant proper motion anomaly -- i.e. difference between the proper motion measured by either \textit{Gaia} or \textit{Hipparcos} and the long-term \textit{Hipparcos}-\textit{Gaia} proper motion -- from \citet{Kervella19_PMa}. In the second column, a P or C indicates that a star is in the planet host or control sample respectively. The third column shows the tangential velocity anomaly, and the fourth column is the minimum companion mass required to explain this anomaly, which we calculated using equation~(15) of \citet{Kervella19_PMa}. In the final column we indicate whether each star has a known binary companion (from Tables \ref{tab:planetsample} and \ref{tab:controlsample}), which could therefore be responsible for the anomaly. Only the first ten lines of the table are reproduced here; a full machine-readable version is available online.}
\label{tab:anomalymass}
\end{table}

Of the 152 stars with significant proper motion anomalies, 97 have a known binary companion (where \textit{known} means that a companion is listed in either WDS or \citealt{Rodriguez15_BinaryDebris} and is therefore in Table \ref{tab:planetsample} or \ref{tab:controlsample}), providing a likely explanation for the anomalies of these systems. Of the 55 which do \textit{not} have a known binary companion, 22 are in the planet host sample and 33 are in the control sample. Eight of these `planet hosts' have a proper motion anomaly which places their minimum companion mass above 13~$M_{\mathrm{J}}$, i.e. out of the planetary regime (though all of the minimum masses are below $\sim$50~$M_{\mathrm{J}}$, meaning that they could be brown dwarfs and are not necessarily stars). Three of them -- HD~72659, HD~10647 (q$^1$~Eri) and HD~82943 -- host debris discs, bringing the detection rate to $38^{+17}_{-12}\%$. This value is higher than the detection rate of the full sample (see section~\ref{sec:discproperties}), which is the opposite of what may be expected given that these systems are candidate binaries (\citealt{Rodriguez12_BinaryDebris}; \citealt{Rodriguez15_BinaryDebris}; \citealt{Yelverton19_BinaryDebris}). However, given the very small numbers involved, it is clear that cutting out the eight stars with potential high-mass companions will not appreciably change the detection statistics of the planet host sample. 

For completeness, we reran all statistical tests on samples of Sun-like stars with no close binaries, with the additional exclusion of the eight planet hosts with astrometrically inferred >13~$M_{\mathrm{J}}$ companions discussed above, and of all 33 control stars with unexplained significant proper motion anomalies. The resulting $p$-values were close to 0.4, similar to those in the bottom row of Table~\ref{tab:pvalues}. Thus, to summarise, using astrometry to obtain a companion mass estimate (following \citealt{Kervella19_PMa}) which is complementary to that from RV only rules out eight of the planet candidates as being genuine planets, though the candidate masses we obtained were lower limits, and there may be more low-inclination stellar (or brown dwarf) companions remaining. Our conclusions from section~\ref{sec:allplanets} are thus unchanged. Note that a more thorough analysis of proper motion anomalies -- for example, using the separations and masses of known binary companions to check whether the known companions can in fact be responsible for the measured anomalies, rather than simply discarding all known binaries -- would be possible, but is beyond the scope of this paper. However, based on our brief investigation here, there is no evidence that such an analysis would strengthen the planet-disc correlation.

\subsection{Temperature comparison}
\label{sec:tempcomparison}

\begin{figure}
	\centering
    \hspace{-0.5cm}
	\includegraphics[width=0.5\textwidth]{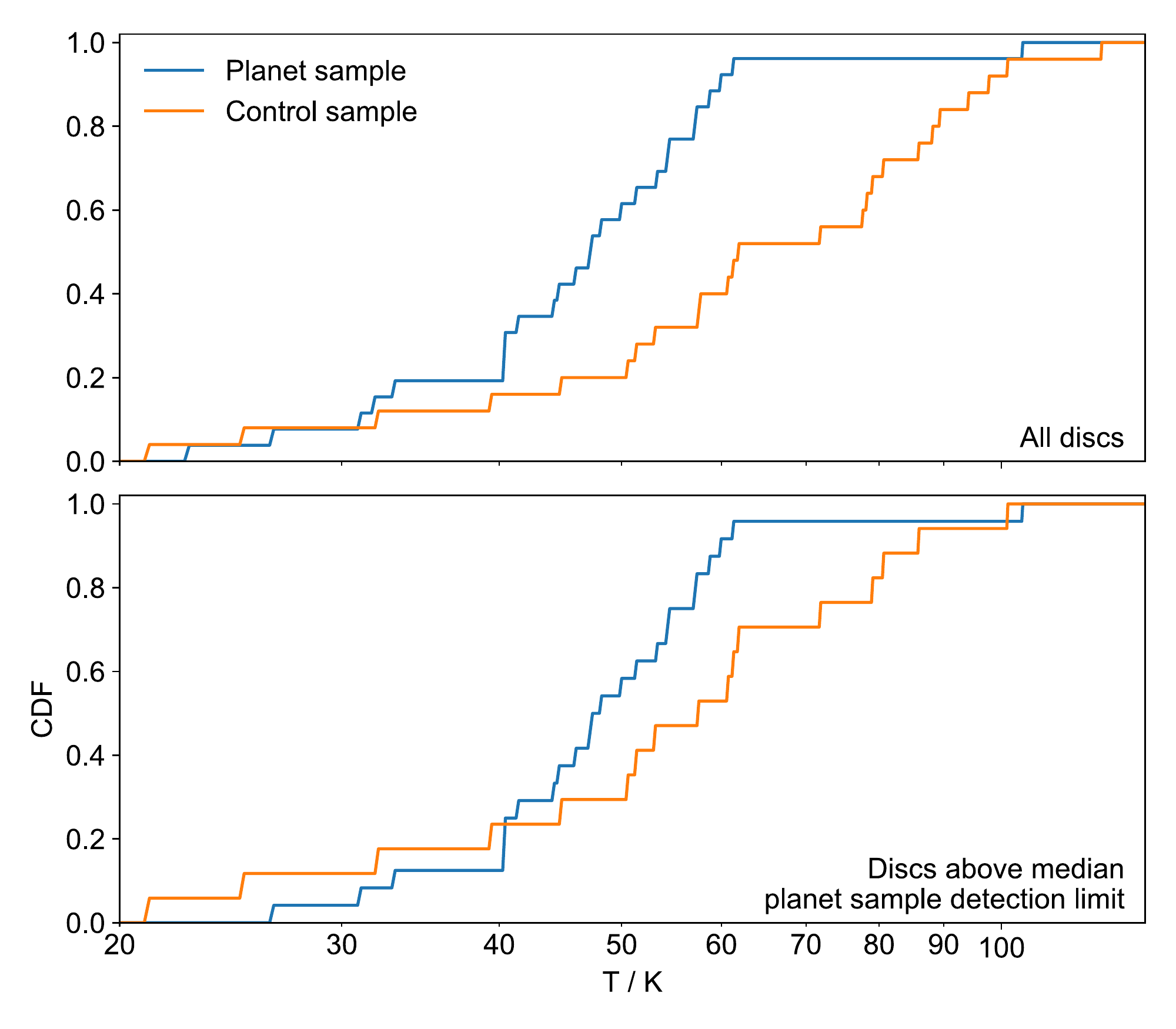}
	\caption{Cumulative distribution functions of black body dust temperatures for systems with a disc detection. The upper panel includes all discs, while the lower includes only systems whose discs lie above the detection threshold for the median planet host sample star shown in Fig.~\ref{fig:fvsT}. The discs around the control stars tend to be warmer, but the difference between the temperature distributions of the two samples becomes less significant once the comparison is restricted to only discs that could have been detected around a typical star in the planet host sample.} 
	\label{fig:temp_cdfs}
\end{figure}

Having established that there is no strong evidence for a link between the presence of planets and the fractional luminosities of debris discs, we proceed to compare the dust temperatures of the two samples, since temperature is the other fundamental parameter that characterises discs. The upper panel of Fig.~\ref{fig:temp_cdfs} shows CDFs of the black body temperatures of the discs in each sample. Note that these are simply conventional CDFs, and differ from the distributions shown in Figs~\ref{fig:KM_estimators} and \ref{fig:KM_estimators_mass} in that they do not include any information from the non-detections. We do not use survival analysis since we cannot place limits on the temperatures of undetected discs in the same way as for the fractional luminosities. 

A Kolmogorov-Smirnov (KS) test applied to these distributions returns a $p$-value of 0.0011, indicating that the temperatures of the two populations are different, with $\sim$3$\sigma$ confidence. The discs we detect around the control stars tend to be warmer than those around the planet hosts. However, this difference may be a result of the different dust sensitivities of the two samples. Recall that while the sensitivity varies between individual systems, the control sample generally has better sensitivity. From Fig.~\ref{fig:fvsT} it can be observed that the warmest discs in the control sample are also the faintest. This correlation can be explained through collisional evolution: warmer discs are closer to their star, so that their planetesimals are moving with higher relative velocities, and thus they decay more quickly through destructive collisions (\citealt{Sibthorpe18_DEBRIS}). The median sensitivity curves in Fig.~\ref{fig:fvsT} illustrate that the population of relatively warm, faint discs found in the control sample is situated in a region of parameter space which is inaccessible for a typical star in the planet sample. Recall also from section~\ref{sec:sample_properties} that the planet hosts are likely somewhat older on average than the control stars, and thus the warmest discs of the planet hosts may have decayed to lower fractional luminosities than those of the control stars.

For a fairer comparison between the samples, we next consider subsamples of only those systems whose discs lie above the typical planet sensitivity curve (i.e. the blue line in Fig.~\ref{fig:fvsT}). The resulting CDFs are shown in the lower panel of Fig.~\ref{fig:temp_cdfs}. While the control sample discs still tend to be warmer than those of the planet hosts, the KS $p$-value increases to 0.0735 (i.e. less than 2$\sigma$ confidence) following the sensitivity cut. Therefore, we conclude that once the detection limits are taken into account, the presence of planets cannot be considered to have a significant influence on the temperatures of debris discs. This is not surprising, given that the planets in our sample are mostly within a few au, while the discs we detect are typically at tens of au. 

\subsection{Implications}
\label{sec:implications}

In this subsection, we first discuss what the fact that we have not found evidence for a significant correlation between the presence of planets and the properties of debris discs may imply about the formation and evolution of planets. We then consider how our results may be interpreted in light of the possible correlation between debris discs and directly imaged planets identified by \citet{Meshkat17_DIPlanets}.

\subsubsection{Planet formation}
\label{sec:pltform}

The Disk Substructures at High Angular Resolution Project (DSHARP; \citealt{Andrews18_DSHARP}) was a survey of 20 protoplanetary discs using the Atacama Large Millimeter/submillimeter Array (ALMA). This survey revealed that gaps and rings are ubiquitous in protoplanetary discs (see e.g. Fig.~3 of \citealt{Andrews18_DSHARP}). One way in which these substructures can be explained is through the presence of planets with masses between Neptune and Jupiter orbiting at tens of au (\citealt{Zhang18_DSHARPPlanets}). As this population of planets is currently inaccessible to other methods of planet detection, and the DSHARP targets are young (with typical ages of $\sim$1~Myr), it could either be the case that the inferred DSHARP planets are independent of the population of confirmed Neptune to Jupiter mass planets at $\sim$1~au like those in our sample, or that the planets in our sample formed at tens of au (like the DSHARP planets), then migrated in to their present-day locations (\citealt{Lodato19_NewbornPlts}). The latter scenario would suggest a direct connection between the planets in our sample and the debris discs we detect, which can be thought of as remnants of the protoplanetary disc phase and are also typically at tens of au. Thus, the lack of a significant planet-disc correlation may point towards the idea that the currently known RV and transiting planets do not form at tens of au, and are independent from the DSHARP planets. Alternatively, it could be the case that the close-in planets in our sample do form at tens of au, but migrate in at an early stage when the planetesimals which will go on to form Kuiper belt analogues have not yet formed. We note that if the DSHARP planets \textit{do} represent an early-stage view of planets like those in our sample, then this could explain why the debris discs of our planet hosts tend to be cooler than those of the control sample (though recall from section~\ref{sec:tempcomparison} that this is not a significant effect), since during their migration from relatively large distances to closer than a few au the planets would clear out circumstellar material from the warmest part of the system.

It is also interesting to contrast the apparent lack of correlation (or anti-correlation) between debris discs and close-in giant planets with the previously established result that discs are around half as common in binary systems closer than $\sim$20~au as they are around single stars or stars with wide stellar companions (\citealt{Yelverton19_BinaryDebris}). There are competing theories for the formation of both giant planets and close binaries. The formation of close binaries may result from turbulent fragmentation of a molecular cloud, or may occur somewhat later via gravitational instability in a circumstellar disc (see e.g. \citealt{Kratter11_CloseBinaries} for a review). Giant planets could either form via core accretion or, like close binaries, via the direct gravitational collapse of a circumstellar disc (see e.g. \citealt{Mordasini10_PltFormation} for a review). Thus, it is possible that giant planets and close binary companions form via essentially the same mechanism. If this were the case, then it may be expected that giant planets and close binaries have similar effects on debris discs, which is not what we observe. This could suggest that the two kinds of companion do \textit{not} both form in the same way, though it could alternatively be the case that they do both form via gravitational instability but that the more massive stellar companions stir planetesimal discs more effectively and thus drive faster collisional decay.

\subsubsection{Comparison with directly imaged planets}
\label{sec:DI}

\citet{Meshkat17_DIPlanets} found tentative evidence that giant planets at tens to hundreds of au -- i.e. those probed by direct imaging -- are more common in systems with detected debris discs than in those without, at the 88\% confidence level. While this result cannot be considered statistically significant, it is nonetheless suggestive that planets at separations comparable to typical debris discs have a stronger correlation with these discs than the much closer-in planets studied in this paper (recall from Table~\ref{tab:pvalues} that we found a difference at a lower confidence level of $\sim$70\%). The apparently contrasting results for planets on close and wide orbits could be understood in terms of a simple model in which the majority of planets form via core accretion regardless of separation, and in which RV planets like those in our study form within a few au (rather than migrating in from large semi-major axes). Within this model, detection of a planet through RV would suggest that planetesimals were able to form in the inner planetary system, while detection of a Kuiper belt analogue would indicate that planetesimals formed in the outer region of the system. However, it is not necessarily the case that the formation of planetesimals at close separations implies that the same should be possible at much larger separations (and vice versa), since, for example, the dust density in protoplanetary discs tends to decrease with stellocentric distance (e.g. \citealt{Andrews2009_PPDisks}), while ice particles that may aid planetesimal growth cannot exist within a few au of the star (e.g. \citealt{Johansen14_Planetesimals}). Additionally, the dynamical environment in the inner and outer systems may be different if undetected perturbing planets at intermediate separations are present, so that even if planetesimals formed across a wide range of separations, they may be quickly ejected from some regions but not others, in a way that depends on the architecture of the system. Thus, it may be expected that the correlation between debris discs and RV planets is weak. In contrast, directly imaged planets will have formed and experienced any subsequent dynamical evolution at separations comparable with typical debris discs, so that a stronger correlation may be expected.

Since core accretion may struggle to produce giant planets at large separations (where the density of primordial circumstellar material is typically low and orbital time-scales are long; e.g. \citealt{Lissauer87_AccretionTimescales}), it is perhaps more plausible that directly imaged planets typically form via gravitational collapse. If this is the case, it is less clear why such planets would be found preferentially in systems with detected planetesimal belts. One possibility relates to stirring, which refers to the excitement of eccentricities (and hence relative velocities) of planetesimals in a disc to a level that causes collisions to become destructive, thus initiating dust production via a collisional cascade and rendering the disc observable. The required stirring may be provided by perturbations from planets, as explored by \citet{Mustill09_PlanetStirring}. From their equation~(15), a debris disc of radius 100~au (which is typical of what is inferred from resolved images; see Fig.~3 of \citealt{Hughes18_Review}) could be stirred by a 5$M_{\mathrm{J}}$ planet of eccentricity 0.1 at 50~au in $\sim$1 Myr, while a planet with the same mass and eccentricity at 2~au (which, from Fig.~\ref{fig:plt_properties}, is typical of the giant planets in our sample) would have a stirring time-scale of $\sim$10~Gyr, around the lifetime of a Sun-like star. If planetary perturbations represent the dominant stirring mechanism in observed debris discs, and the systems we studied do not usually have additional massive planetary companions on wide orbits, then it could be the case that the probability of hosting a debris disc is the same for systems with RV and directly imaged planets, but that a larger fraction of the discs in RV planet systems remain unstirred and therefore undetected.

Finally, we note that while it is possible to speculate on the interpretation of our findings juxtaposed with those of \citet{Meshkat17_DIPlanets}, differences in the ages of the systems studied in our work mean that our results are not directly comparable. The systems studied by \citet{Meshkat17_DIPlanets} have typical ages below a few hundred Myr, a result of bias towards young systems in direct imaging surveys. In contrast, our control sample was not selected based on age, while our planet host stars are in fact biased towards older ages (as discussed in section~\ref{sec:sample_properties}). Thus, while we have not compiled age information, we expect that our systems are largely older than a Gyr, given the typical lifetimes of stars with spectral types later than mid-F. It may thus be the case that any planet-disc correlation that exists for young stars does not persist into old age, since debris discs with the same radii and planetesimal properties but with different initial masses (and hence fractional luminosities) will tend towards the same mass at late times through collisional evolution (\citealt{Wyatt07_Transience}).

\section{Conclusions}
\label{sec:conclusions}

We compiled a sample of 201 stars known to host planets (mostly detected through RV) and a control sample of 294 stars without known planets. As detailed in section~\ref{sec:samples}, the samples are constructed purely of stars with existing far infrared observations from \textit{Spitzer} or \textit{Herschel}, so that we can identify cool infrared excesses indicating the presence of Kuiper belt-like debris discs. Comparing the two samples in section~\ref{sec:sample_properties}, we identified some important differences in their properties: the planet hosts are more distant (and thus generally have poorer sensitivity to dust), and the control sample has a much greater proportion of close binaries and of late-type stars, which is likely a result of target selection biases in RV planet surveys.

In section~\ref{sec:SED_modelling} we described our SED modelling procedure, in which a model consisting of a star and optionally one or two modified black body components representing dust emission is fit to each system's SED. We used the results of this modelling to define a criterion for an excess to be considered significant, as well as to derive the fractional luminosities $f$ and temperatures $T$ of the discs we identify. We presented these disc parameters in section~\ref{sec:discproperties}, showing that the discs we detect typically have temperatures of $\sim$50~K and fractional luminosities of a few $10^{-5}$. 

We proceeded to compare the fractional luminosities of the two samples in section~\ref{sec:allplanets}, using survival analysis to take account of the samples' different typical sensitivities. Comparing the full samples using statistical tests suggested that their fractional luminosity distributions differ with $\sim$3$\sigma$ confidence. However, guided by the fact that close binaries and late-type stars are both known to have a negative impact on disc detection rates, we then cut out systems with close (<~135~au) binary companions and restricted the comparison to Sun-like stars (with temperatures between 4700 and 6300~K), which reduced the significance of the difference to $\sim$1$\sigma$. This led us to our main conclusion: there is no evidence that the presence of RV planets significantly affects the fractional luminosities of debris discs.

We also compared the fractional luminosities of low-mass and high-mass planetary systems, with the distinction between the two subsamples being the presence or absence of a planet more massive than Saturn. In section~\ref{sec:planetmass} we showed that there was no evidence for a difference between the $f$ distributions of these subsamples, with survival analysis giving $p$-values greater than 0.9. This contrasts with \citet{Wyatt12_LowMassPlts}, who found a tentative result that debris discs are more often detectable in low-mass planetary systems, but we agree with their conclusion about the presence of a disc for most of the stars they considered, suggesting that their result was simply due to their much smaller sample sizes. 

In section~\ref{sec:stellarcompanions} we briefly considered (following \citealt{Kervella19_PMa}) how \textit{Gaia} astrometry may be used to identify planet candidates which are really low-inclination stellar companions and thus have a small $M\sin(i)$ value. We found that only eight of the 201 planet hosts' companions can be ruled out as planetary-mass objects based on an initial analysis of the stars' astrometry. The results of the survival analysis tests are almost unchanged following a cut of systems whose proper motion anomaly suggests the presence of a companion with a mass in the brown dwarf or stellar regime, since the number of systems this cuts out is small. 

Then, in section~\ref{sec:tempcomparison} we compared the temperature distributions of the planet hosts and the control stars, finding that the planet hosts tend to have cooler discs, with a KS test on the temperatures returning a $p$-value of 0.001, indicating a $\sim$3$\sigma$ significant difference. However, we concluded that this difference is probably a result of the different sensitivities of the two samples. The warmest discs in the control sample are also the faintest -- perhaps a result of their faster collisional evolution -- and would therefore be impossible to detect around a typical star in the planet host sample (since the planet hosts are further away). Considering only discs which would be detectable around a typical planet host star, we found a $p$-value of 0.07, reducing the significance to below $2\sigma$.

Finally, in section~\ref{sec:implications} we considered the implications of our findings. In particular, we discussed how the lack of a significant planet-disc correlation may suggest that the currently known RV and transiting planets do not originate from the inward migration of the young Neptune to Jupiter mass planets at tens of au which have recently been inferred from gaps in protoplanetary discs. We also discussed the contrast between our result and the tentative correlation between debris discs and planets at large separations identified by \citet{Meshkat17_DIPlanets}. We suggested that the correlation may be stronger for such planets than for the close-in planets in our sample because they are located at similar semi-major axes to -- and therefore likely shared common formation conditions and dynamical histories with -- typical debris discs, and/or because they stir the discs on short time-scales while discs in systems with only close-in planets have not yet had a collisional cascade initiated.

Our overall conclusion is thus that there is currently no evidence for a significant difference in the fractional luminosities or temperatures of debris discs around stars with and without known planets, even when taking advantage of data from far infrared surveys that specifically targeted planet hosts (in particular, the \texttt{OT1_gbryden_1} \textit{Herschel} survey). Our findings are consistent with those of \citet{MoroMartin15_PlanetsAndKBs}, who also did not find a significant correlation between discs and planets of either low or high mass. However, since their samples were based on \textit{Herschel} surveys whose targets were selected irrespective of planet presence, the number of planet hosts in their work was small, totalling 22. Thus, based on their work it was unclear whether the apparent lack of a correlation was a genuine physical effect or an artefact of the small numbers involved. Our Sun-like planet host sample with close binaries excluded is considerably larger, containing 138 systems; we have thus demonstrated that any planet-disc correlation must indeed be weak.

In the absence of a new instrument observing at $\sim$100~$\mu$m, the clearest way to make further progress on the problem of planet-disc correlation would be to assemble a control sample of systems with RV planet non-detections, either by analysing archival RV data or by performing a new RV survey of stars with existing far infrared observations. Additionally, \textit{Gaia} will ultimately lead to the discovery and publication of a new population of astrometric planets and binaries; considering that \textit{Gaia} is an all-sky survey, the newly discovered planet hosts will have some overlap with past far infrared surveys, thus permitting a study similar to that performed in this paper, and allowing further exploration of how disc detection rate depends on companion mass as discussed in section \ref{sec:implications}.

\section*{Acknowledgements}

We thank the reviewer for a report that helped to improve the quality of this paper. BY acknowledges the support of an STFC studentship, and is grateful to the Institute of Astronomy at the University of Cambridge for additional funding. GMK is supported by the Royal Society as a Royal Society University Research Fellow. 




\bibliographystyle{mnras}
\bibliography{refs}





\bsp	
\label{lastpage}
\end{document}